\documentclass[a4paper,fleqn]{cas-dc}
\usepackage[numbers,sort&compress]{natbib}
\usepackage{caption}
\usepackage{physics}
\usepackage{siunitx}
\usepackage{float}
\usepackage{subfig}
\usepackage{blindtext}

\newcommand{\czts}{Cu$_2$ZnSnS$_4$}
\newcommand{\cztsse}{Cu$_2$ZnSn(S,Se)$_4$}
\newcommand{\czgs}{Cu$_2$ZnGeS$_4$}
\newcommand{\czss}{Cu$_2$ZnSiS$_4$}
\newcommand{\cigs}{CuInGa(S,Se)$_2$}

\newcommand{\cfr}{\textit{cfr.}}
\newcommand{\etal}{\textit{et.al.}}
\newcommand{\voc}{$V_{\mathrm{OC}}$}
\newcommand{\jsc}{$J_{\mathrm{SC}}$}
\newcommand{\ai}{\textit{ab initio}}
\newcommand{\ie}{i.\,e.,}
\newcommand{\angstrom}{\mbox{\normalfont\AA}}

\def\tsc#1{\csdef{#1}{\textsc{\lowercase{#1}}\xspace}}
\tsc{WGM}
\tsc{QE}
\tsc{EP}
\tsc{PMS}
\tsc{BEC}
\tsc{DE}

\DeclareUnicodeCharacter{2009}{\,} 

\begin{document}
\let\WriteBookmarks\relax

\shorttitle{Opto-electronic properties and solar cell efficiency modelling of Cu$_2$ZnXS$_4$ (X=Sn,Ge,Si) kesterites}
\shortauthors{T Ratz \etal}

\captionsetup[figure]{labelfont={bf},name={Figure},labelsep=period, justification={justified}}
\captionsetup[table]{labelfont={bf},name={Table},labelsep=period, justification=justified}

\title [mode = title]{Opto-electronic properties and solar cell efficiency modelling of Cu$_2$ZnXS$_4$ (X=Sn,Ge,Si) kesterites}                      

\author[1,2]{Thomas Ratz}[orcid=0000-0002-3629-1087]

\cormark[1]
\fnmark[1]
\ead{thomas.ratz@uliege.be}
\credit{Conceptualization, Methodology, Formal analysis, Investigation, Visualisation, Writing - original draft}

\author[1,3]{Jean-Yves Raty}[orcid=0000-0001-7535-2834]
\credit{Supervision, Validation, Ressources, Writing - original draft}
\author[4]{Guy Brammertz}[orcid=0000-0003-1404-7339]
\credit{Supervision, Project administration, Validation, Writing - original draft}
\author[2,4,5]{Bart Vermang}[orcid=0000-0003-2669-2087]
\credit{Supervision, Project administration, Validation, Writing - original draft}
\author[1]{Ngoc Duy Nguyen}[orcid=0000-0002-0142-1611]
\credit{Supervision, Project administration, Validation, Writing - original draft}

\address[1]{CESAM | Q-MAT | Solid State Physics, Interfaces and Nanostructures, Physics Institute B5a, Allée du Six Août 19, B-4000 Liège, Belgium}
\address[2]{Institute for Material Research (IMO), Hasselt University, Agoralaan gebouw H, B-3590 Diepenbeek, Belgium}
\address[3]{University of Grenoble Alpes | CEA-LETI | MINATEC Campus | Rue des Martyrs 17, F-38054 Cedex 9 Grenobles, France}
\address[4]{IMEC division IMOMEC | partner in Solliance, Wetenschapspark 1, B-3590 Diepenbeek, Belgium}
\address[5]{Energyville, Thor Park 8320, B-3600 Genk, Belgium}

\begin{abstract}
In this work, first principle calculations of \czts \ (CZTS), \czgs \ (CZGS) \ and \czss \ (CZSS) are performed to highlight the impact of the cationic substitution on the structural, electronic and optical properties of kesterite compounds. Direct bandgaps are reported with values of 1.32, 1.89 and 3.06 eV respectively for CZTS, CZGS and CZSS. 
In addition, absorption coefficient values of the order of 10$^4$ cm$^{-1}$ are obtained, indicating the applicability of these materials as absorber layer for solar cell applications. In the second part of this study, \ai \ results (absorption coefficient, refractive index and reflectivity) are used as input data to model the electrical power conversion efficiency of kesterite-based solar cell. In that perspective, we used an improved version of the Shockley-Queisser theoretical model including non-radiative recombination via an external parameter defined as the internal quantum efficiency. Based on predicted optimal absorber layer thicknesses, the variation of the solar cell maximal efficiency is studied as a function of the non-radiative recombination rate. Maximal efficiencies of 25.88 \%, 19.94 \% and 3.11 \% are reported respectively for \czts, \czgs \ and \czss \ for vanishing non-radiative recombination rate. 
Using a realistic internal quantum efficiency which provides \voc \ values comparable to experimental measurements, solar cell efficiencies of 15.88, 14.98 and 2.66 \% are reported respectively for \czts, \czgs \ and \czss \ (for an optimal thickness of 1.15 $\mu$m). 
With this methodology we confirm the suitability of \czts \ in single junction solar cells, with a possible efficiency improvement of 10\% enabled through the reduction of the non-radiative recombination rate. In addition, \czgs \ appears to be an interesting candidate as top cell absorber layer for tandem approaches whereas \czss \ might be interesting for transparent photovoltaic windows.
\end{abstract}

\begin{highlights}
\item[-] Insights of Sn substitution by Ge and Si in S-kesterite compounds
\item[-] Connection between \ai \ predicted material properties and device characteristics
\item[-] Quantitative correlation between cell efficiency and non-radiative recombination rate
\item[-] CZTS confirmed for single junction cell with a 10\% possible efficiency increase
\item[-] CZGS uses as top cell in tandem approach whereas CZSS could be used for PV windows
\end{highlights}


\maketitle


\section{Introduction}
\label{Introduction}

Over the years, photovoltaic (PV) thin film technology has emerged as an interesting candidate for efficient and large-scale energy production. To this aim, this technology must fulfill several criteria such as low-cost thin film synthesis, high solar cell efficiency and materials resources availability and accessibility \cite{Vesborg:2012gt}. In relation with the latter point, the European Commission has identified Ga and In as critical raw materials and  highlighted the scarcity of those elements used for the synthesis of inorganic chalcogenide \cigs \ (CIGS) alloys implemented as absorber layer for PV applications \cite{EUCRM}. Despite the high efficiency reported for CIGS solar cells, with a record value of 23.3$\%$ \cite{Green:2020ga, nakamura2019cd}, the incorporation of this material in a large-scale energy production technology might be compromised. This justifies an urgent search for alternative compositions with comparable or better efficiencies than CIGS. As a consequence, over the past 20 years, the scientific community has been investigating kesterite \cztsse \ materials as absorber layer in solar cell applications \cite{Giraldo:2019iia}. Benefiting from the well-established knowledge of CIGS, kesterite-based solar cell efficiency gradually increased over the years, reaching values of 12.6$\%$ for \cztsse \ \cite{Wang:2013gs} and 11$\%$ for \czts \ \cite{Yan:2018dw} using various chemical \cite{Todorov:2020ir} or physical \cite{Ratz:2019cs} routes for the synthesis of the kesterite thin films. 

However, new challenges concerning further efficiency improvements have recently arisen. Large open circuit voltage \voc \ deficits have been reported as responsible for the efficiency limitation encountered \cite{Grossberg:2019gt, Giraldo:2019iia}. Several elements have been pointed out as possible culprits for the \voc \ deficits, including interface recombination due to bands misalignment \cite{PlatzerBjorkman:2019ed, crovetto2017band}, formation of secondary phases, and/or high intrinsic point defect concentration leading to non-radiative recombination in the kesterite bulk material \cite{Grossberg:2019gt}. As a result, recombination centres are present both at the architectural level (band misalignments with the buffer layer) and at the compositional/morphological level (intrinsic point defects or secondary phases) within the absorber layer \cite{Giraldo:2019iia}. Focusing on the kesterite absorber layer, several solution paths have been considered to overcome the current efficiency limitation, like alloying using isoelectronic substitution elements such as Ag for Cu, Ge for Sn or Se for S \cite{Romanyuk:2019cq, Li:2018du} or via the cationic substitution of Zn or Sn \cite{kumar2018substitution, tablero2014electronic}.

In the past, alternative kesterite materials have been studied both theoretically and experimentally, leading to promising efficiencies  for Ge-containing kesterite compounds \cite{Kim:2016jr, Giraldo:2018cg, Buffiere:2015dd, Choubrac:2018ex, vermang2019wide,khelifi219path}. Using density functional theory (DFT) calculations, a few works reported predictions over structural properties, electrical properties or optical properties of alternative kesterite materials such as \czts \ \cite{Chen:2010gk, Khare:2012gj, Zamulko:2017cy, Liu:2012dea}, \czgs \ \cite{Chen:2010gk, Shu:2013ed, Zamulko:2017cy, Liu:2012dea, vermang2019wide} and \czss \ \cite{Chen:2010gk, Shu:2013ed, Zamulko:2017cy, Liu:2012dea}. However, the variety of computational approaches do not facilitate the comparison of the materials physical properties. In addition, to the best of our knowledge, the DFT results are rarely compared to experimental measurements.

In this work, we first investigate theoretically the cationic substitution of Sn by two other iso-electronic elements: Ge and Si, in kesterite \czts . The structural and opto-electronic properties are calculated for \czts \ as the reference material \cite{Kim:2018jd,Chen:2013cna}, \czgs \ as a promising material regarding the experimental efficiency achieved \cite{Ratz:2019cs, vermang2019wide} and \czss \ as an interesting candidate regarding the elemental abundance \cite{Vesborg:2012gt}. Then, the obtained \ai \ results are used as input data to feed an improved version of the Shockley-Queisser model, allowing us to connect the intrinsic material properties to the solar cell macroscopic properties. Via this cell efficiency modelling, physical quantities such as the open circuit voltage \voc, the short circuit current density $J_{SC}$ and the fill factor $FF$ are computed.

In the first section of this paper, the structural properties of the materials are presented. Then, in the following sections, the Heyd–Scuseria–Ernzerhof exchange-correlation functionnal (HSE06) \cite{heyd2003hybrid} is used to compute the electronic and optical properties. Based on the band structures and the densities of states (DOS), the electrical properties of the materials are reported and compared. To complete the investigation, the optical properties are presented and related to the electrical ones. This approach allows us to extract the general trends highlighting the impact of the cationic substitution of Sn by Ge and Si on the opto-electronic properties. In the second part of this work, using the \ai \ results as input data, the upper limit of the kesterite-based solar cell efficiency is calculated using the theoretical model proposed by Blank \etal \ \cite{blank2017selection}. This model allows us to compute physical quantities that can be compared to experimental results such as the solar cell efficiency $\eta$ using as parameters the solar cell temperature $T$, the absorber layer thickness $d$ and the internal quantum efficiency $Q_i$ \cite{blank2017selection}.

\section{Computational method}

First principle calculations have been performed using Vienna \textit{Ab initio} Simulation Package (VASP) code \cite{kresse1996efficiency} with the Projector-Augmented Wave (PAW) potential method \cite{kresse1999ultrasoft}. Perdew-Burke-Ernzerhof (PBE) GGA pseudo-potentials \cite{perdew1996generalized} were used with orbitals Sn 4d and Ge 3d treated as valence electrons. Ionic and electronic relaxation were achieved using a cut-off energy of 550 eV and a $\Gamma$-centered uniform \textbf{k}-points mesh of $6 \times 6 \times 6$ \textbf{k}-points. Applying the strongly constrained and appropriately normed semilocal density functional (SCAN) \cite{sun2015strongly, sun2016accurate}, the structures were relaxed until the numerical convergence regarding the self-consistent cycles reaches forces between ions less than $10^{-4}$ eV/$\SI{}{\angstrom}$. The system total energy was converged down to $10^{-6}$ eV. During relaxation, the symmetry was kept constant to the kesterite point group symmetry ($I-4$) and the atomic positions, cell volume and cell shape were allowed to relax. Starting from the relaxed structure, the Heyd–Scuseria–Ernzerhof exchange-correlation functionnal (HSE06) \cite{heyd2003hybrid} known for its bandgap prediction accuracy \cite{heyd2005energy}, was used to compute the electronic and optical properties. 

\section{Results and discussion}
 
\subsection{Structural properties}
 
\begin{figure}[!h]
\includegraphics[width=\columnwidth]{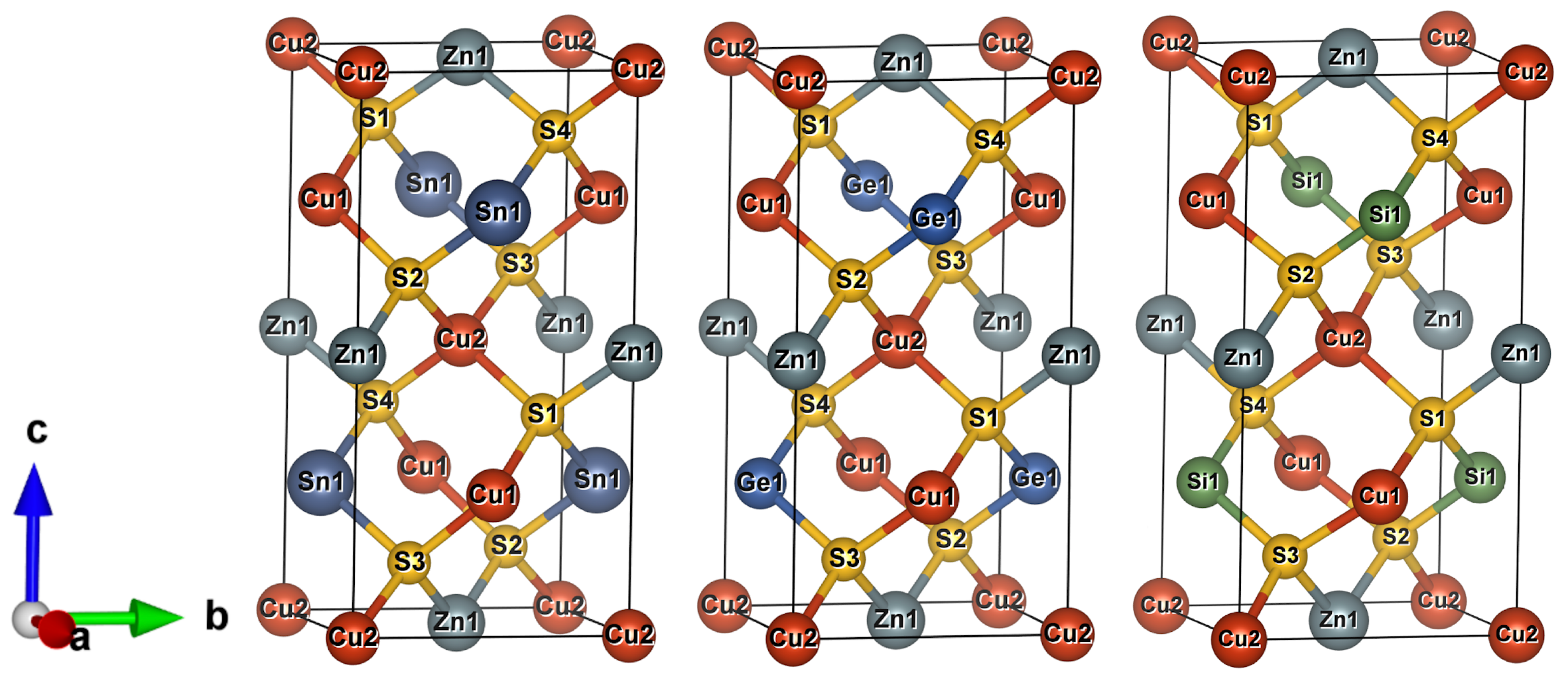}
\caption{Representation of the conventional cells of the Cu$_2$ZnXS$_4$ (X=Sn,Ge,Si) kesterites.}
\label{Structural_param}
\end{figure}

The lattice parameters $a$, $b$, $c$ (\cfr \ Fig.\ref{Structural_param}), the conventional cell volume $V$ and the atomic distances $d_{\mathrm{Cu-S}}$ and $d_{\mathrm{X-S}}$ (X=Sn,Ge,Si) were obtained as a result of the ionic relaxation (Table \ref{Structural_param}). 

The sequential substitution of Sn by Ge and Si induces a contraction of the kesterite lattice parameters. A reduction of $a$ and $b$ from 5.40 \AA \ (\czts) to 5.25 \AA \ (\czss) is observed while the $c$ parameter is reduced from 10.79 \AA \ to 10.32 \AA . The results reported in Table \ref{Structural_param} are in good agreement with experimental measurements for \czts \ \cite{guo2009synthesis, levcenko2012free, lisunov2013features, Zamulko:2017cy, Dun:2014dt, walsh2012kesterite} and \czgs \ \cite{khadka2013study}. To our knowledge, experimental characterisation of Si-pure kesterite crystal structures has not been reported yet. According to Refs. \cite{hamdi2014crystal, levcenco2011polarization}, an orthorhombic crystalline structure is observed for high Si concentrations. Nevertheless, several theoretical works reported values close to $a,b = 5.25$ \AA \ and $c = 10.32$ \AA \  \cite{Zamulko:2017cy, Liu:2012dea} as reported here. This lattice contraction can be interpreted by taking into account the successive reduction of the atomic radius of the substitutional cation from $r_{Sn} = 1.45$ $\SI{}{\angstrom}$, to $r_{Ge} = 1.25$ $\SI{}{\angstrom}$ and to $r_{Si} = 1.10$ $\SI{}{\angstrom}$ \cite{slater1964atomic}. In addition, the cationic substitution implies a successive reduction of the distances $d_{\mathrm{X-S}}$ between the cation and the sulphur atom, highlighting the variation of the equilibrium distances between the atoms as a result of the change in bond ionicity. Consequently, the conventional cell volume decreases from 314.9 $\SI{}{\angstrom}^3$ for the Sn-containing compound to 294.87 $\SI{}{\angstrom}^3$ for \czgs \ and to 283.94 $\SI{}{\angstrom}^3$ for \czss. One can also notice that the cation substitution does not impact the $d_{\mathrm{Cu-S}}$ distances. In the following section, the results presented here will be put into perspective with the electronic properties. 

\begin{table}[!h]
\centering
\resizebox{\columnwidth}{!}{%
{\renewcommand{\arraystretch}{1.5}
\begin{tabular}{ c   c  c  c   c  c  c  c }
\hline
Materials & a,b [\AA] & c [\AA] & V [\AA$^3$] & $d_{\mathrm{X-S}}$  [\AA]  & $d_{\mathrm{Cu-S}}$  [\AA] & Exp. & Theo.\\
\hline
\hline
   Cu$_2$ZnSnS$_4$ &    5.40     &  10.79  &  314.90  &   2.44   &  2.29 & \cite{guo2009synthesis, levcenko2012free, lisunov2013features}  & \cite{Zamulko:2017cy, Dun:2014dt, walsh2012kesterite, Liu:2012dea}   \\ 
   
   Cu$_2$ZnGeS$_4$&    5.30     &  10.51   &  294.87  &  2.26  & 2.28  &  \cite{khadka2013study} & \cite{Zamulko:2017cy, Liu:2012dea, zhang2012structural} \\ 
   
   Cu$_2$ZnSiS$_4$&    5.25     &  10.32  &  283.94  &  2.15  & 2.28 & \cite{hamdi2014crystal} & \cite{Zamulko:2017cy, Liu:2012dea}\\ 
\hline
\end{tabular}%
}
}
\caption{Lattice parameters $a,b$ and $c$ (see \ Fig. \ref{Structural_param}) and conventional cell volume $V$ of Cu$_2$ZnXS$_4$ (X=Sn,Ge,Si) kesterites. Interatomic distances between the cation (X=Sn,Ge,Si)  and the sulphur atom $d_{\mathrm{X-S}}$ are reported as well as the copper-sulphur distances $d_{\mathrm{Cu-S}}$.}
\label{Sparam}
\end{table}

\begin{center}
\begin{table*}[!tp]
\centering
\resizebox{0.7\textwidth}{!}{%
{\renewcommand{\arraystretch}{1.5}
\begin{tabular}{ c  c  c  c  c  c  c  c  c  c  c  c }
\hline
\multirow{2}{*}{Materials}  & \multirow{2}{*}{$E_G$ [eV] (Exp.)} & \multicolumn{4}{c}{$m_{//}^*$ [$m_e$]}        & \multicolumn{4}{c}{$m_{\perp}^*$ [$m_e$]}  & \multirow{2}{*}{$\epsilon_{\infty}$  [$\epsilon_0$]} & \multirow{2}{*}{Theo.}  \\ 
                                                         &                     & $\Gamma_{v,1}$ & $\Gamma_{v,2}$ & $\Gamma_{v,3}$ & $\Gamma_{c}$    & $\Gamma_{v,1}$ & $\Gamma_{v,2}$ & $\Gamma_{v,3}$ & $\Gamma_{c}$&  &   \\ 
                                    \hline
                                     \hline
 Cu$_2$ZnSnS$_4$  &  1.32 (1.50 \cite{Ratz:2019cs})  &  -0.69  &   -3.32   &   -0.16    &   0.19     &   -0.77  &   -0.64    &  -0.19  &	   0.18   &  6.77   & \cite{Liu:2012dea, Chen:2010gk, Zamulko:2017cy} \\
 
 Cu$_2$ZnGeS$_4$  &  1.89 (1.90  \cite{khadka2013study}) &  -0.72   &  -3.49   &   -0.19   &   0.23   &  -0.72   &  -0.63  &   -0.24  &   0.22   &  6.44 & \cite{Liu:2012dea, Chen:2010gk, Zamulko:2017cy} \\

  Cu$_2$ZnSiS$_4$  &  3.06 (N.A.)  &   -1.44     &  -3.65     &  -0.25     &    0.26   &  -1.63   &    -0.68    &   -0.33   &   0.25  & 5.78 & \cite{Liu:2012dea, Chen:2010gk, Zamulko:2017cy} \\
\hline
\end{tabular}%
}
}
\caption{Bandgaps $E_G$ and effective masses $m^*$ scaled by the free electron mass $m_0$ of Cu$_2$ZnXS$_4$ (X=Sn,Ge,Si) kesterites. Effective masses have been calculated around the $\Gamma$ high symmetry \textbf{k}-point and along two directions in the reciprocal space: [0,0,0] to [0,0,1] (resp. [0,0,0] to [0,1,0]) for the first effective mass component $m_{\perp}$ (resp. for the second component $m_{//}$). High-frequency dielectric constants $\epsilon_\infty$ of the materials are also presented and scaled with the vacuum electrical permittivity $\epsilon_0$.}
\label{electronic_param}
\end{table*}
\end{center}

\subsection{Electronic properties}

\begin{figure}[!h]
\centering
\subfloat[Cu$_2$ZnSnS$_4$]{\includegraphics[height=0.25\textheight]{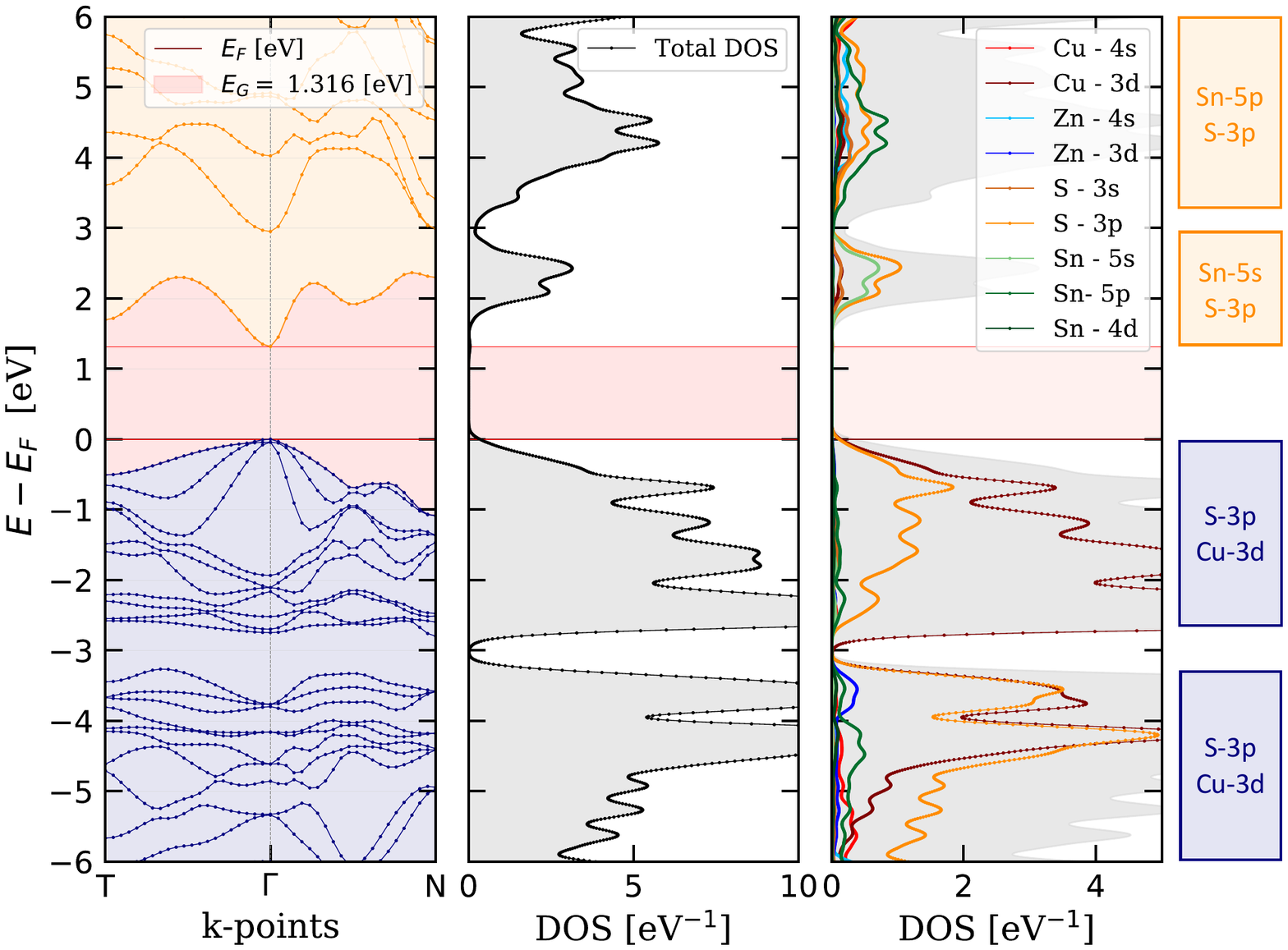}\label{Bands_CZTS}}\\
\subfloat[Cu$_2$ZnGeS$_4$]{\includegraphics[height=0.25\textheight]{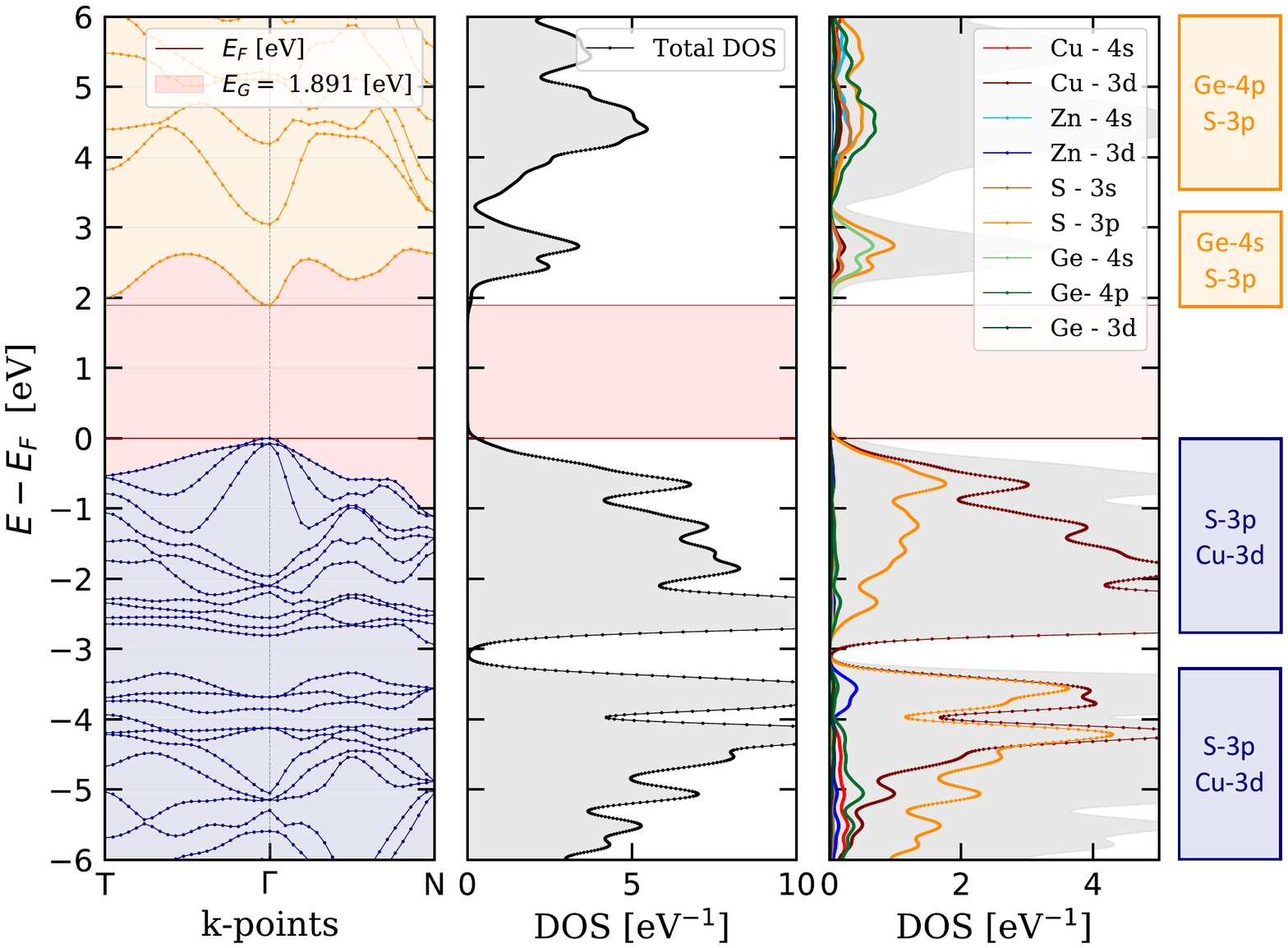}\label{Bands_CZGS}}\\
\subfloat[Cu$_2$ZnSiS$_4$]{\includegraphics[height=0.25\textheight]{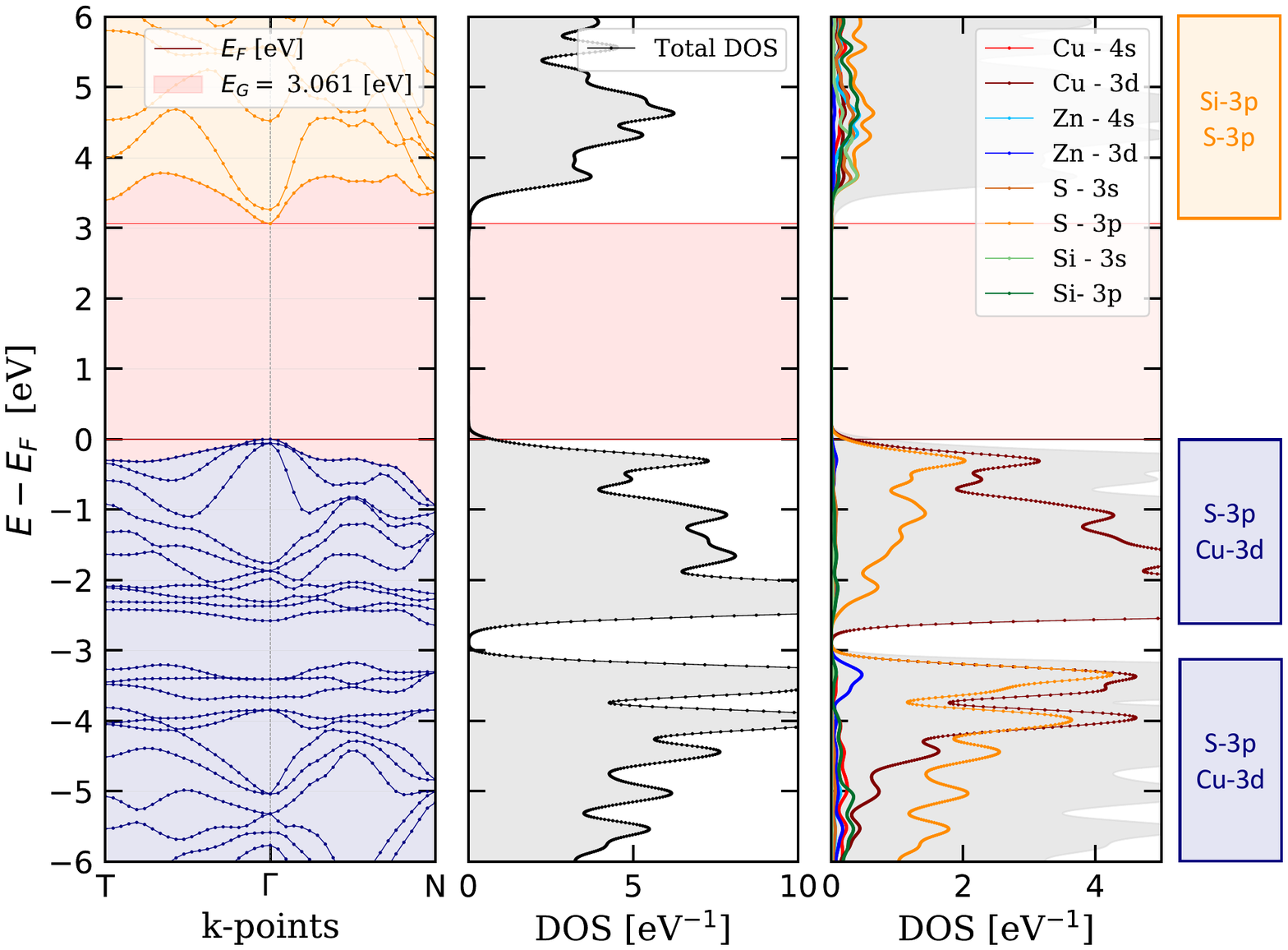}\label{Bands_CZSS}}
\caption{Band structures, densities of states and orbital projected densities of states of Cu$_2$ZnXS$_4$ (X=Sn,Ge,Si) kesterites. The densities of states are presented with an applied gaussian smearing of 0.08 eV. The band dispersion is calculated along $T$: [0,0,1/2] - $\Gamma$: [0,0,0] - $N$: [1/2,1/2,1/2]. Main atomic orbital contributions to the DOS are presented alongside the figures.}
\label{Bands}
\end{figure}

As it can be observed in Fig. \ref{Bands}, all calculated kesterite bands present a direct bandgap located at the $\Gamma$ point. The bandgap energy $E_G$ increases from 1.32 eV for \czts \ to 1.89 eV  for \czgs \ and to 3.06 eV for \czss \ as reported in Table \ref{electronic_param}. These results are comparable to those reported by Zamulko \etal \ in their theoretical investigation \cite{Zamulko:2017cy}. In comparison to experimental values, the Sn-containing kesterite bandgap is underestimated by 0.18 eV as usual reported values are around 1.5 eV \cite{Ratz:2019cs}. In contrast, the \czgs \ bandgap value of 1.89 eV fits with the reported experimental bandgaps of 1.88 and 1.93 eV \cite{khadka2013study}. According to Ref. \cite{vishwakarma2018structural}, a bandgap value of 2.71 eV was experimentally obtained for \czss.

We provide here a focus on the orbitals projected DOS and their contributions to electronic states in the band structure, for the Sn-kesterite compound (Fig. \ref{Bands_CZTS}). The main contributions to the conduction band states come from S 3p and Sn 5s atomic orbitals close to the bottom of the band and S 3p and Sn 5p atomic orbitals for higher energy levels. Concerning the valence band, the hybridisation between Cu 3d and S 3p orbitals provide the main contributions to energy states at the top of the band \cite{paier2009cu}. This tendency is also observed for the two other kesterite materials, \textit{i.e.} the bottom of the conduction band is formed by either the s atomic orbital of the cation X (X=Sn, Ge) or the p orbital of the cation Si and the 3p orbital of the chalcogen S, while the contributions to the top of the valence band come from the 3d atomic orbital of Cu and the 3p atomic orbital of the sulphur element. 

For \czgs \ and \czss, the substitution of Sn by Ge and Si (Figs. \ref{Bands_CZGS} \& \ref{Bands_CZSS}) seems to slightly flatten the energy level at the bottom of the conduction band. 
The bandgap increase from 1.32 to 3.06 eV is due to the variation of the chemical interaction between the cation and the sulphur, which leads to (i) a weak flattening of the energy level at the bottom of the conduction band and (ii) a shift of this energy level towards higher energies. To link those observations to the structural properties of the materials one can put into perspective the decrease of the cation/sulphur interatomic distance $d_{\mathrm{X-S}}$ with the change in the chemical bonding and the increase of the kesterite bandgap. In contrast, the substitution of the cation atoms leaves the valence band unchanged as the orbitals contributing to these states are from Cu and S for which the interatomic distances $d_{\mathrm{Cu-S}}$ are reported constant from one kesterite material to another (\cfr \ Table \ref{Structural_param}).

In addition to the bandgaps, the effective masses are presented in Table \ref{electronic_param}. These have been calculated around the $\Gamma$ point, at the direct bandgap location, and along two directions in the reciprocal space: [0,0,0] to [0,0,1] for the first effective mass component $m_{\perp}$ and along [0,0,0] to [0,1,0] for the second component $m_{//}$. As shown in Fig. \ref{Bands}, one energy level is present at the bottom of the conduction band and three energy levels are located at the top of the valence band. Consequently, the effective masses have been calculated for the lowest energy level in the conduction band named $\Gamma_{c}$ and for the three highest energy levels at the top of the valence band $\Gamma_{v,1}$, $\Gamma_{v,2}$, $\Gamma_{v,3}$, labeled from the highest energy level to the lowest one. For both the conduction and valence band, the general trend observed is a slight increase of the effective mass absolute value when Sn is sequentially substituted by Ge and Si. Then, as kesterite materials behave electrically as p-type semiconductor \cite{Grossberg:2019gt}, we first discuss the hole effective mass values. As presented in Table \ref{electronic_param}, concerning the $m^*_{//}$ component, $\Gamma_{v,2}$ effective masses are significantly higher than $\Gamma_{v,1}$ and $\Gamma_{v,3}$, highlighting the presence of light and heavy holes in this particular direction. In addition, similar values are reported regarding $\Gamma_{v,1}$ and $\Gamma_{v,3}$ for the perpendicular component while, in contrast, $\Gamma_{v,2}$ is one order of magnitude lower than in the parallel direction. Concerning the electron effective masses, similar values are obtained for both components $m^*_{//}$ and $m^*_{\perp}$ with a slight increase from a minimal value of 0.18 $m_e$ to a maximal value of 0.26 $m_e$ observed as the Sn cation is substituted. Those results are in good agreement with those obtained by Liu \etal \ with reported effective masses of 0.18, 0.21 and 0.26 $m_e$ \cite{Liu:2012dea}. This suggests that the hole and electron effective masses would only slightly increase as Sn is substituted by Ge and then Si.
In summary, the cationic substitution does not impact significantly the hole nor electron effective masses but leads to a significant increase of the kesterite bandgap.

\subsection{Optical properties}
\label{optical_section}

Following the computation of the electronic properties, the optical properties of the kesterite materials have been determined via the calculation of the dielectric tensor $\epsilon (E)$ whose real $\epsilon_1$ and imaginary $\epsilon_2$ parts are shown in Fig. \ref{dielectric} (see supplementary material for the detailed equations). In this figure, the components $xx$, $yy$ and $zz$ of $\epsilon (E)$ are presented for each compound. It appears that the sequential substitution of Sn with Ge and Si leads to a decrease of the high frequency dielectric response $\epsilon_\infty$ from 6.77 $\epsilon_0$ (\czts) to 6.44 $\epsilon_0$ (\czgs) and reaching 5.78 $\epsilon_0$ for the Si-containing compound (\cfr \ Table \ref{electronic_param}). As expected, the decrease in $\epsilon_\infty$ is in agreement with the increase of the materials bandgap. Concerning the imaginary part of the dielectric tensor $\epsilon_2 (E)$, the onset of absorption is also shifted towards higher energies as the bandgap increases.

Then, the absorption coefficient $\alpha(E)$ as well as the reflectivity $R(E)$ and refractive index $n(E)$ are computed as described in the supplementary material. In Fig. \ref{absorption} the absorption coefficient of the materials are presented alongside the solar irradiance spectrum. First, one can notice that each compound exhibits an absorption coefficient of the order of 10$^{4}$ cm$^{-1}$ within the energy range of non-negligible solar irradiance (between 0.5 and 4 eV). This result highlights the applicability of these kesterite materials as absorber layer in solar cell applications. However, an energetic shift of the absorption curves is also observed from the Sn-containing kesterite to the Si-containing kesterite with a first absorption peak located at the respective bandgap energies of the materials. The \czts \ and \czgs \ curves have a similar behaviour while for the Si-containing kesterite curve, the plateau observed for the two other kesterites disappears as a consequence of the energy level shift at the bottom of the conduction (\cfr \ Fig. \ref{Bands_CZSS}). Finally, in Fig. \ref{reflectionindex}, the refractive index $n(E)$ and reflectivity $R(E)$ are presented. As reported, the refractive indices at 0 eV are 2.59, 2.53 and 2.40 respectively for \czts, \czgs \ and \czss \ with variations of 0.6 in values between 0 and 5 eV. Concerning the reflectivity values, a variation from 20 to 30$\%$ within the 0 to 5 eV energy range is observed. Additionally, it is worth noticing some reflectivity differences of nearly 10\% between \czts \ and \czss \ for some energy values. 

\begin{figure}[!h]
\centering
\subfloat[Dielectric tensor]{\includegraphics[height=0.25\textheight]{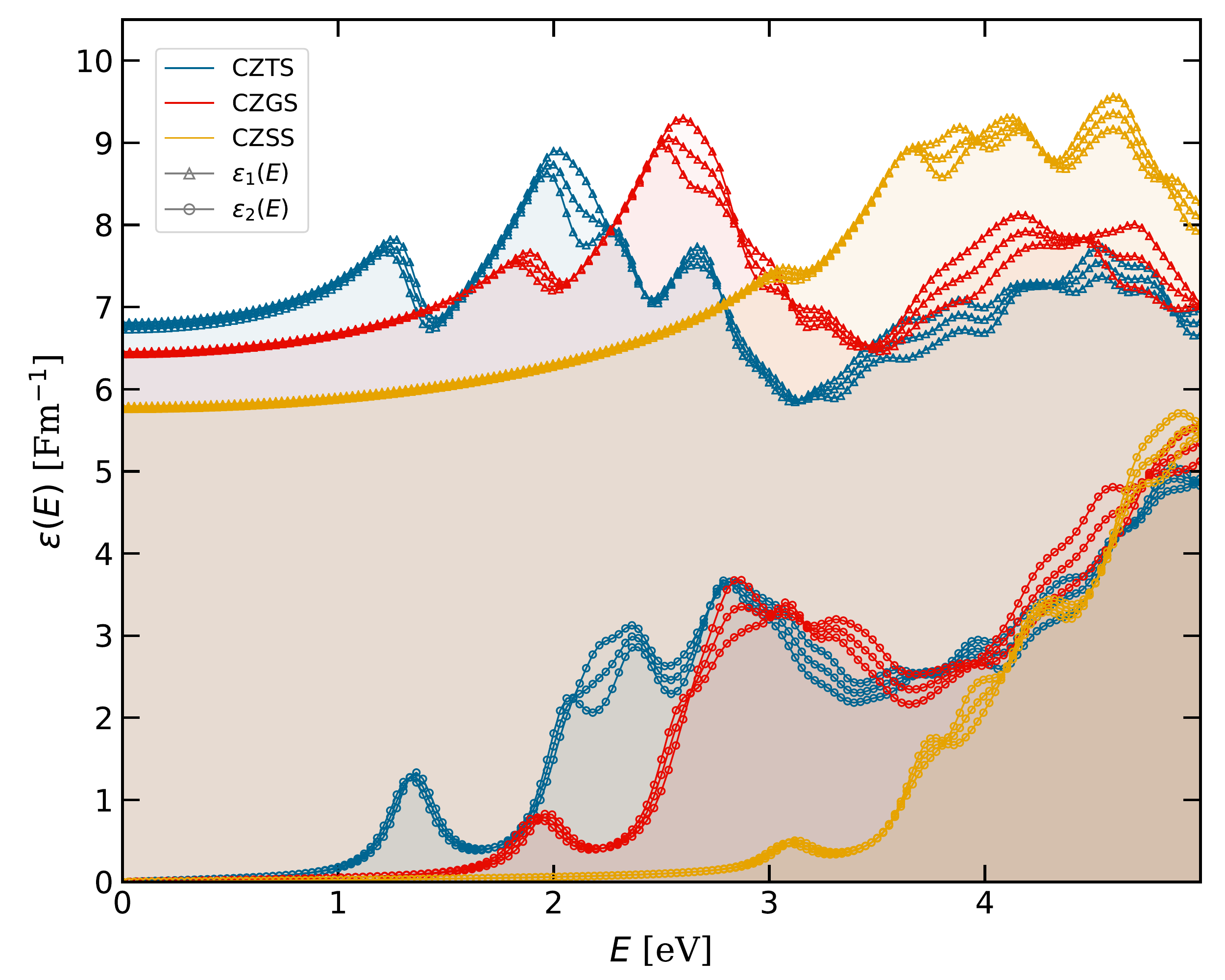}\label{dielectric}} \\
\subfloat[Absorption coefficient]{\includegraphics[height=0.25\textheight]{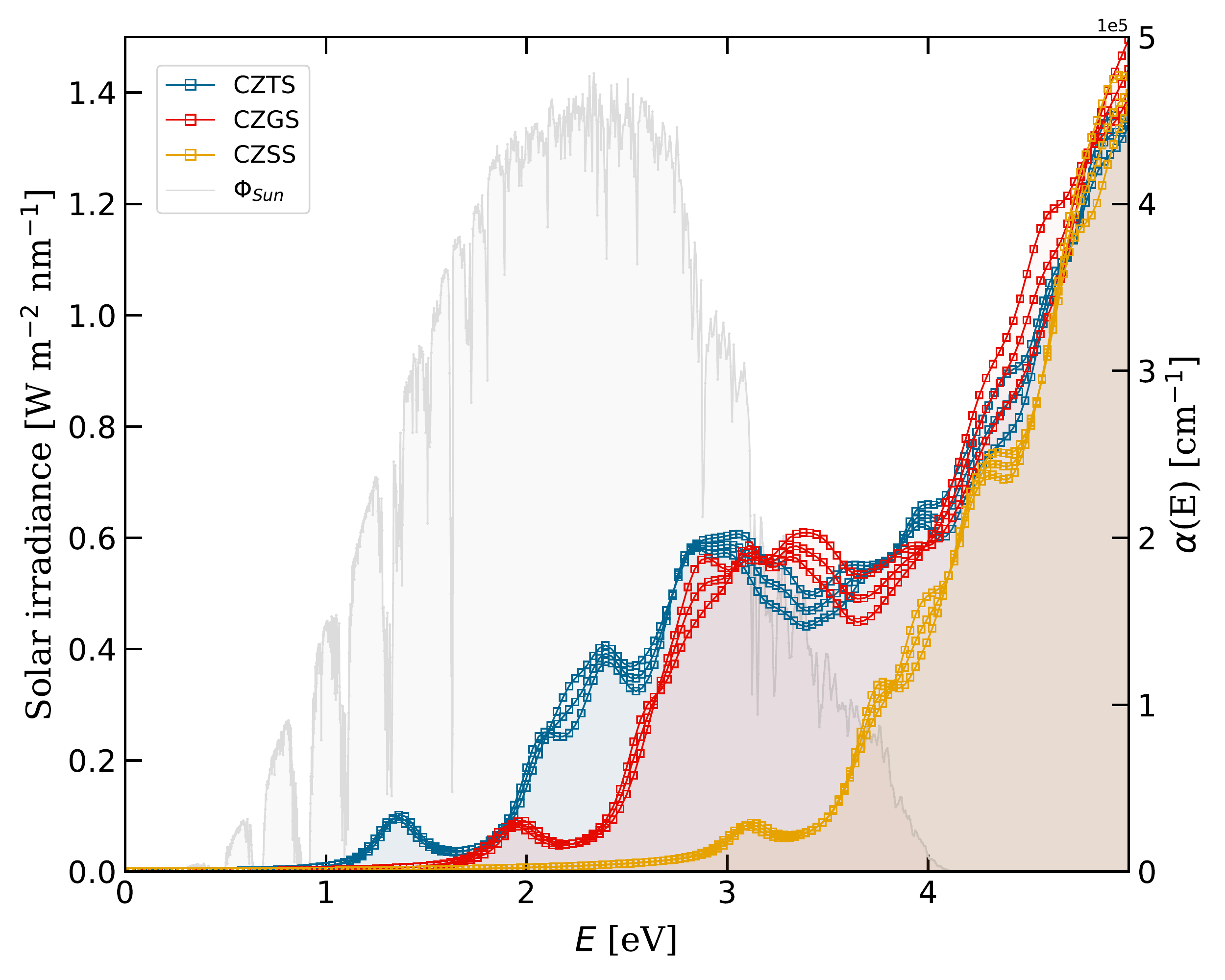}\label{absorption}} \\
\subfloat[Refractive index and reflectivity]{\includegraphics[height=0.25\textheight]{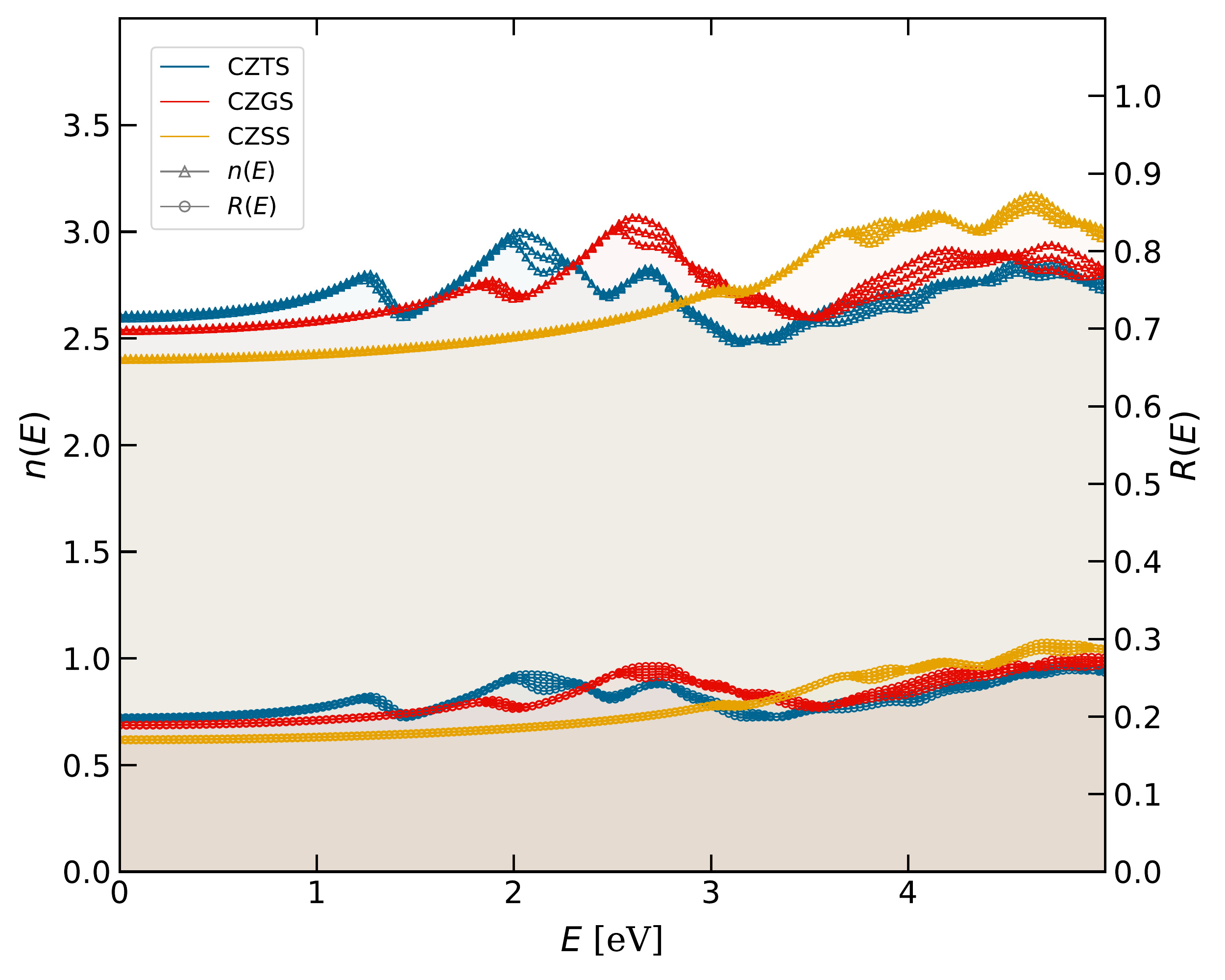}\label{reflectionindex}}
\caption[font=small,justification=justified]{(a) Real $\epsilon_1$ and imaginary $\epsilon_2$ parts of the dielectric tensor $\epsilon(E)$. For each compound, the $xx$, $yy$ and $zz$ components of the tensor are presented. (b) The absorption coefficients $\alpha(E)$ and the solar irradiance spectrum are presented. (c) Materials refractive indices $n(E)$ and reflectivity $R(E)$ spectra.}
\label{Optic}
\end{figure}

\subsection{Electrical power conversion efficiency}

In this section, we focus on the theoretical modelling of solar cell macroscopic physical quantities such as the short circuit current density \jsc, the open circuit voltage \voc \ and the solar cell electrical power conversion efficiency $\eta$ using Cu$_2$ZnXS$_4$ (X=Sn,Ge,Si) as absorber layer. The predictions are realised based on the theoretical model presented by Blank \etal \ \cite{blank2017selection}. The improvements proposed by Blank \etal \ over the Shockley-Queisser model are (i) the use of the internal quantum efficiency $Q_i$ as a model parameter to take into account non-radiative recombinations and (ii) the incorporation of light trapping by taking into account the refractive index $n(E)$ in the calculation of the radiative current density $J_{\mathrm{rad},0}(n,d)$ \cite{blank2017selection} (see supplementary material). 

Non-radiative recombinations occur via defects (intrinsic point defect, defect cluster or grain boundary) in the bulk material acting as recombination centres which impact the solar cell properties. Therefore, in this theoretical work, we chose to use this physical quantity as a model parameter. In that perspective, the internal quantum efficiency is expressed as the ratio between the radiative recombination rate $R_{\mathrm{rad},0}$ and the total recombination rate: $R_{\mathrm{rad},0} + R_{\mathrm{nrad},0}$, leading to a non-radiative recombination rate under equilibrium conditions,

\begin{equation}
 R_{\mathrm{nrad},0} = R_{\mathrm{rad},0}  \frac{(1-Q_i)}{Q_i}
\label{Rnrad_main}
\end{equation}

Considering a perfectly crystalline material, all recombinations are radiative and the photons emitted (\ie \ not reabsorbed) contribute to the emission spectrum of the material which, in this model, is assumed as the black body spectrum at temperature $T$ = 300 K. These radiative recombinations are therefore thermodynamically required and are proportional to the amount of electrons within the conduction band (\ie \ proportional to the the temperature). This first situation corresponds to an internal luminescence quantum efficiency $Q_i$ value equals to unity for which the total recombination rate $R_{0}$ is equal to the radiative recombination rate $R_{\mathrm{rad},0}$. If one considers intrinsic point defects and defect clusters within the bulk material, the recombinations become of both types: radiative and non-radiative. The thermodynamic condition of emission must still be fulfilled ($R_{\mathrm{rad},0}$) and additionally, recombinations via recombination centres occur in the bulk materials ($R_{\mathrm{nrad},0}$), leading to an increase of the total recombination rate $R_{0}$. In this paper, the $Q_i$ value is related to the amount of non-radiative recombinations within the bulk material which is proportional to the number of radiative recombinations (Eq. (\ref{Rnrad_main})). $Q_i$ can consequently be related to the internal quantum efficiency $IQE$ which is an experimentally measured physical quantity. The detailed description of the theoretical model proposed by Blank \etal \ is presented in the supplementary material. To feed this theoretical model we use the previously calculated optical results ($\alpha(E)$, $n(E)$ and $R(E)$) as input data. It is worth noticing that the computed material properties obtained corresponds to a perfect crystal (\ie \ $Q_i = 1$). As the internal quantum efficiency tends to vanish, variations of the optical properties are expected as defects will introduce new electronic states. However, in this work the perfect crystal optical properties are considered for each value of $Q_i$. Accordingly, the absorptance $A(E)$ of the absorber layer is determined via Eq. (\ref{absorptance_main}), assuming a flat solar cell surface and a thin film thickness $d$:

\begin{equation}
A(E,d) = [1 - R(E)] - \mathrm{exp}(- 2 \alpha(E) d)
\label{absorptance_main}
\end{equation}

\begin{figure}[!h]
\centering
\includegraphics[width=1\columnwidth]{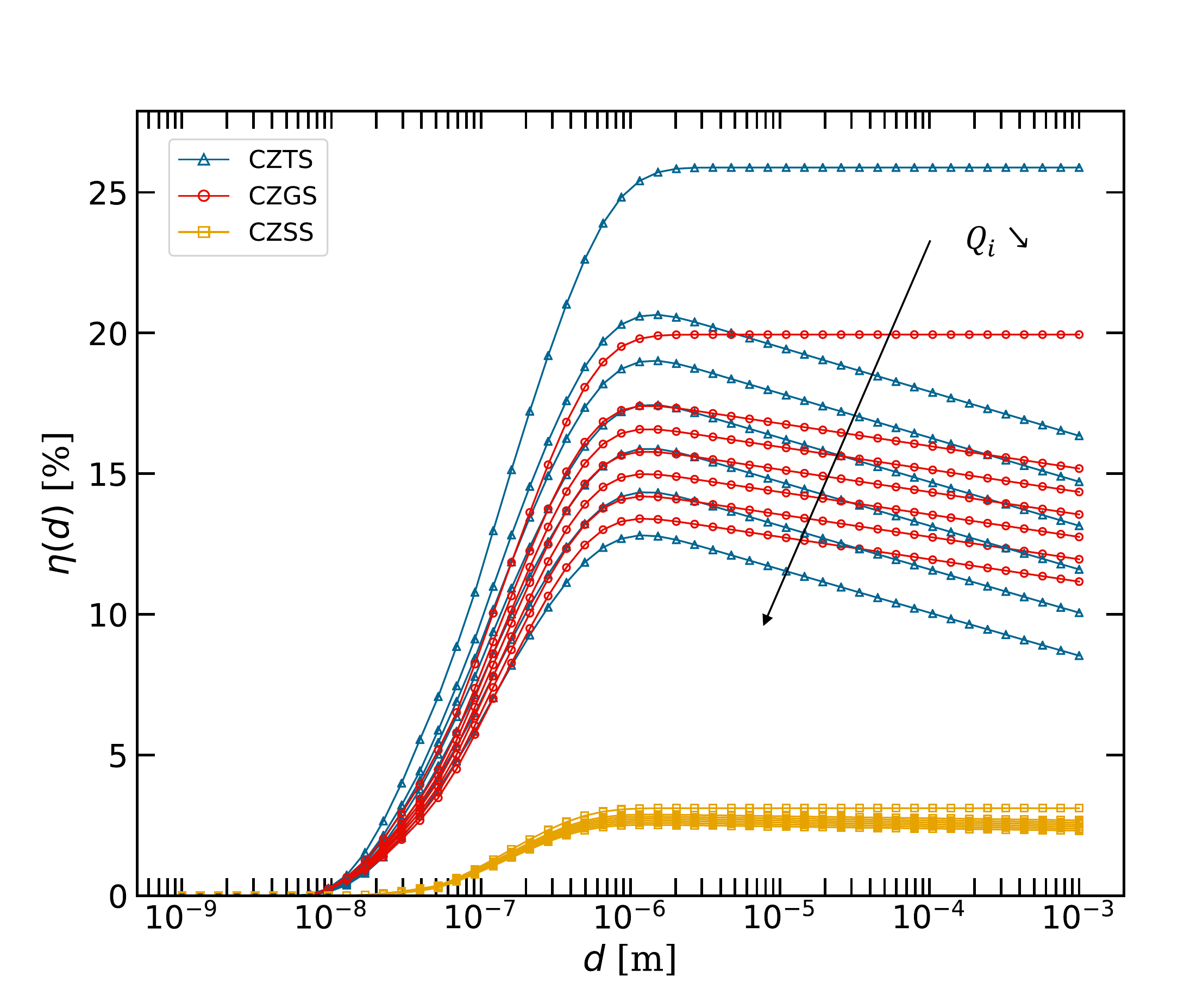}
\caption{Solar cell efficiency modelling presented as a function of the absorber thin film thickness $d$ for various internal quantum efficiency $Q_i \in [10^{-6};1]$.}
\label{EfffigVSthick}
\end{figure}

\begin{figure}[!h]
\centering
\includegraphics[width=1\columnwidth]{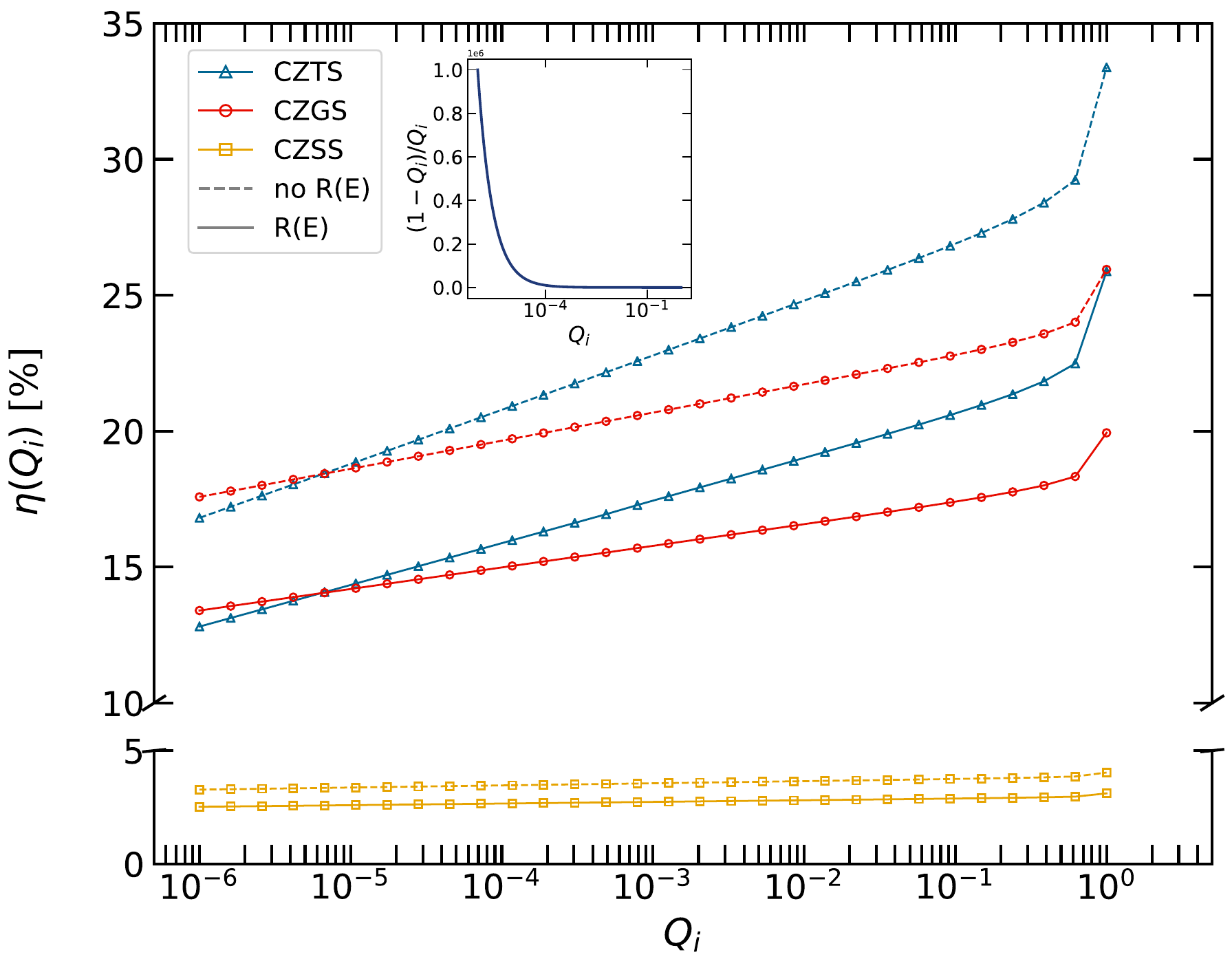}
\caption{Solar cell efficiency modelling for an optimal absorber layer thickness extracted from Fig. \ref{EfffigVSthick} presented as a function of the internal quantum efficiency $Q_i \in [10^{-6};1]$. Results from simulations taking into account the materials reflectivity $R(E)$ are presented in full lines while dashed lines represent the maximal efficiencies obtained assuming $R(E)=0$. In inset, evolution of the prefactor fixing the non-radiative recombination rate as described in Eq. (\ref{Rnrad_main}) with respect of $Q_i$.}
\label{EffVSQifig}
\end{figure}

\begin{figure}[!h]
\centering
\includegraphics[width=0.96\columnwidth]{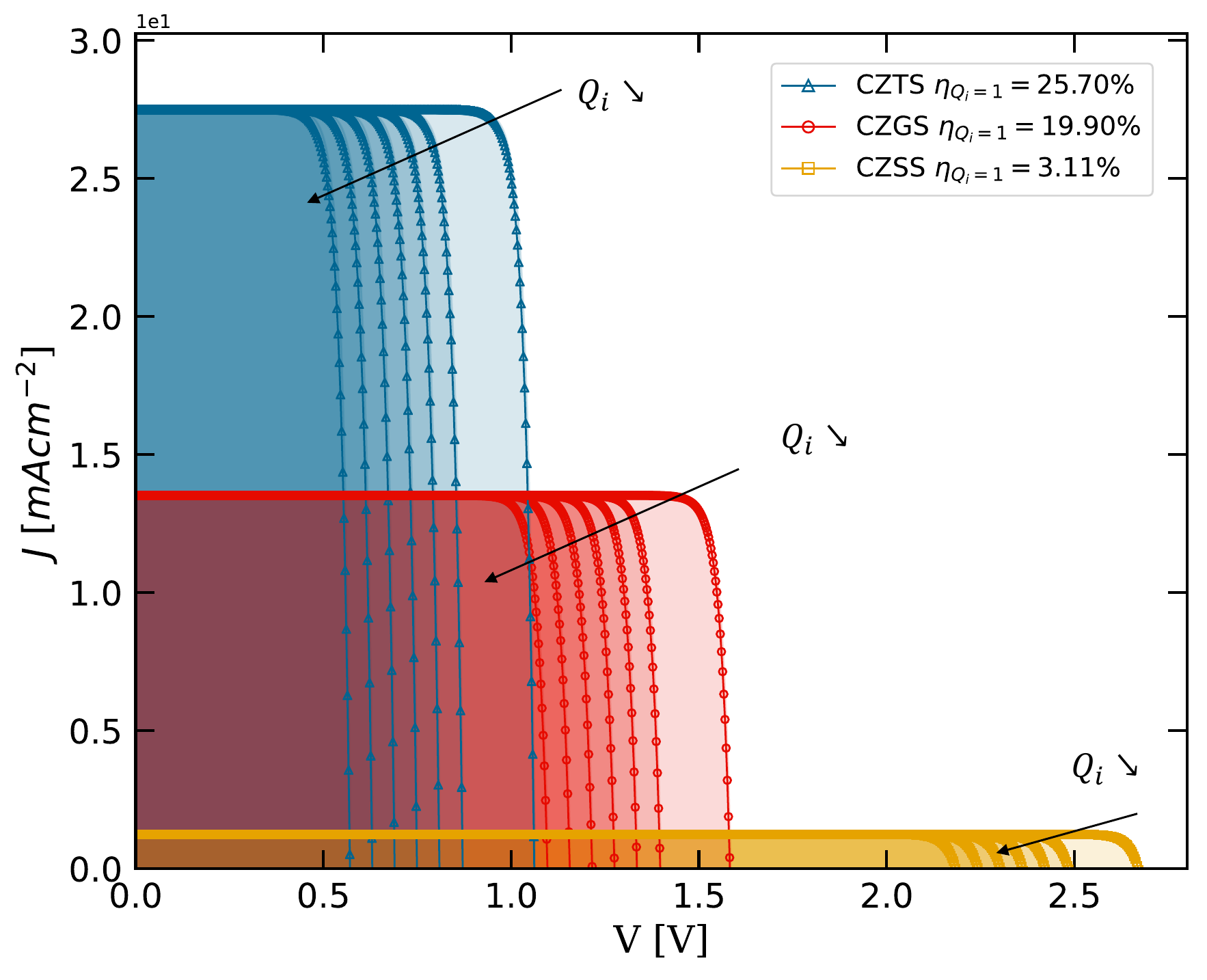}
\caption{Current density-voltage curves of solar cell modelling for various internal quantum efficiency $Q_i \in [10^{-6};1]$. Results obtained for an absorber layer thickness of 1.5 $\mu$m.}
\label{JV}
\end{figure}

The obtained results are presented for a solar cell temperature $T$=300K as follow:
\begin{itemize}
\item[-] First, we evaluate the optimal thicknesses (\ie \ associated to a maximum for $\eta$) of the absorber layer as a function of $Q_i$. To this perspective, the efficiency of the solar cell is calculated for different values of the absorber layer thickness $d$ and for various internal quantum efficiency values $Q_i \in [10^{-6};1]$ (Fig. \ref{EfffigVSthick}).
\item[-] Using this optimal thickness, we compute the maximal efficiency for a range of internal quantum efficiency values $Q_i \in [10^{-6};1]$ (Fig. \ref{EffVSQifig}). In addition, to highlight the impact of the absorber layer reflectivity on the solar cell properties, the calculation is performed with and without taking into account the materials reflectivity $R(E)$ in the calculation of the absorptance $A(E) $ (Eq. (\ref{absorptance_main})).
\item[-] Then, in Fig. \ref{JV}, the current density voltage curves for the respective kesterite-based solar cells are presented for different internal quantum efficiency values $Q_i \in [10^{-6};1]$ and for a usual absorber layer thickness of 1.5 $\mu$m.
\item[-] In Table \ref{efftable}, the main solar cell electrical characteristics are reported first by assuming no non-radiative recombination (\ie \ $Q_i = 1$) and secondly by assuming a non-radiative recombination rate fixed by $Q_i = 10^{-4}$ in order to obtain results comparable to actual experimental device characteristics (\ie \ experimentally comparable \voc \ values). Finally, the results obtained are compared to various experimental works.
\end{itemize}

In Fig. \ref{EfffigVSthick}, the maximal efficiency is calculated as a function of the absorber layer thickness. Each curve represents an internal quantum efficiency value ranging logarithmically from 1 (highest efficiency) to $10^{-6}$ (lowest efficiency). Here, we report a significant disparity between the \czss -based solar cell efficiencies with values below 5\% for all $Q_i$, compared to the cells based on the two other kesterite materials. This observation is linked to the larger bandgap of \czss, which limits drastically the short circuit current density (see supplementary material) as illustrated in Fig. \ref{JV} and Table \ref{efftable}. The general trend observed for all materials is an increase of the efficiency as the absorber thickness increases over 10 nm. Then, for $d$ just below 1 $\mu$m, the efficiency reaches either a plateau (for $Q_i =1$) or a maximal value for an optimal thickness, before decaying linearly as $d$ is increased (for $Q_i < 1$). The optimal thicknesses reported for the absorber layer thin films are between 1.15 and 2.68 $\mu$m (\cfr \ Table \ref{efftable}). The observed increase of $\eta$ with $d$ can be explained by the optimisation of the the absorptance function $A(E)$ which gets closer to $1-R(E)$ for $E > E_G$, thus maximising the short circuit current density. The optimisation of the absorptance also maximises $J_{\mathrm{rad},0}$ which reduces \voc \ and reduces $\eta$ but this phenomenon is not dominant here. Then, for a unit value of $Q_i$, \jsc \ asymptotically reaches a maximum value and any further increase of the thickness (over the optimal thickness value) does not result in any notable increase of the efficiency value. In contrast, for internal quantum efficiency values $Q_i < 1$, as the absorber layer gets thicker, the non-radiative recombination rate increases, leading to a decrease of the open circuit voltage and, consequently, to the efficiency drop (see supplementary material).

From the previous calculations, for each $Q_i$ value, the absorber layer thickness giving the maximum efficiency is extracted as the optimal absorber thickness value $d_{opt}$. Then, in a second calculation (Fig. \ref{EffVSQifig}), the evolution of the maximal efficiency as a function of the internal quantum efficiency for an optimal thickness is reported both without (dashed lines, $R(E)=0$) and with (full lines, $R(E)$ from DFT results in section \ref{optical_section}) taking into account the materials reflectivity in the absorptance calculation (see \ Eq. (\ref{absorptance_main})). Concerning the impact of the materials reflectivity on the solar cell efficiency for the \czts \ compound, depending on the $Q_i$ value, a percentage point loss of 4 to 8 in efficiency is observed  (decrease of 4 to 6 observed for \czgs \ and of 1 for \czss). Concerning the behaviour of $\eta$ with respect to $Q_i$, the cell efficiency increases as $Q_i$ tends to unity and as the non-radiative recombination rate decays towards 0 (see \ Eq. (\ref{Rnrad_main})). Then, as the internal quantum efficiency decreases, the efficiencies reported also decrease with absolute percentage point losses of 1.54, 0.79 and 0.07 per order of magnitude, respectively for \czts, \czgs \ and \czss. The variation of the slopes of the material curves observed in Fig. \ref{EffVSQifig} from one kesterite to another is a direct consequence of the materials optical properties variations. Following the cationic substitution, the variation of the material absorptance function leads to a decrease of the radiative recombination rate value. As a consequence, for lower value of $R_{\mathrm{rad},0}$, a variation of $Q_i$ implies a smaller variation of the non-radiative recombination rate and consequently of the total recombination rate. In addition, any increase of the saturation current density $J_0$ will lead to a decrease of the open circuit voltage and consequently of the efficiency. Combining these two explanations, as the material absorptance gets optimal with respect to the black body spectrum (\ie \ from \czss \ to \czts), the larger the radiative recombination rate is, the larger the efficiency variation per decade of $Q_i$ will be (see supplementary material). Then, following the variation of the slope observed, for a fixed efficiency value (for example 15\%), \czts \ appears more "robust" to larger non-radiative recombination rate as the $Q_i$ value required to reach this efficiency is lower for \czts \ than for \czgs. This highlights the fact that even for a lower ratio of radiative over total recombination rates, a same efficiency is obtained. This tendency is reversed for $Q_i$ value lower than 10$^{-5}$.

As shown in Fig. \ref{JV}, the short circuit current density \jsc \ is independent of $Q_i$ as this one is related to the total number of electron-hole pair (EHP) generated by photons absorption (see supplementary material). This quantity depends only on the absorptance of the materials. Following the cationic substitution, the \jsc \ value decreases as the absorptance function worsen with respect to the solar spectrum. In opposition, an increase of the open circuit voltage is observed as the cation is substituted. Indeed, as the optical properties degrades, the radiative recombination rate decreases and consequently the \voc \ value increases. In addition \voc \ is $Q_i$ dependent. For a given material, as $Q_i$ tends towards a null value, the total recombination rate will increase resulting in a decrease of the \voc \ value, leading to the decrease of the cell efficiency as reported in Fig. \ref{EffVSQifig}. 
Finally, the differences in $\eta$ between the three kesterite materials are associated to the decreasing value of \jsc \ which is not fully compensated by the increase of the \voc \ both attributed to the poorer absorptance as we move from the Sn-containing compound to the Si-containing compound.

\begin{center}
\begin{table*}[!h]
\centering
\resizebox{0.7\textwidth}{!}{%
{\renewcommand{\arraystretch}{1.5}
\begin{tabular}{|c|ccc|ccccc|cc|}
\hline
Materials &      $E_G$ [eV]         &      $R(E)$      &          $Q_i$        & $d_{opt}$ [$\mu$m]         &     \jsc \ [mAcm$^{-2}$]       & $V_{\mathrm{OC}}$ [V]      &       FF [$\%$]     &    $\eta$ [$\%$] & Exp. & Theo. \\ \hline
\hline
\multirow{4}{*}{Cu$_2$ZnSnS$_4$} & \multirow{4}{*}{1.32} & 0 &   1                 &  2.68  &   35.69 &  1.06  & 88.62  &  33.38 & \multirow{4}{*}{\cite{Ratz:2019cs}} & \multirow{4}{*}{\cite{kim2020upper, blank2017selection}}  \\
														&        						     & 0 &   $10^{-4}$   &  1.15  &   35.21   & 0.70 &  84.56  &  20.78  & & \\ 
                     									&        						     & DFT &                 1   &  2.68    &   27.68   & 1.06 &  88.59  &  25.88  & & \\ 
                    					  				&        						     & DFT &   $10^{-4}$   &  \cellcolor{black!10} 1.15    &  \cellcolor{black!10} 27.19   &  \cellcolor{black!10} 0.70 & \cellcolor{black!10}  84.39  & \cellcolor{black!10} 15.88  & & \\ \hline
                    					  				
\multirow{4}{*}{Cu$_2$ZnGeS$_4$} & \multirow{4}{*}{1.89} & 0 &   1                 &   2.02 &   17.62 & 1.58  & 91.65  &  25.95  & \multirow{4}{*}{\cite{vermang2019wide}} & \multirow{4}{*}{\cite{vermang2019wide}}\\
														&        						     & 0 &   $10^{-4}$   &  1.15  &   17.53   &  1.23 & 89.82 &   19.65  & &\\ 
														&        						     & DFT &   1                  &   2.02    &   13.54   &  1.58 & 91.62 &   19.94  & &\\ 
														&        						     & DFT &   $10^{-4}$   &  \cellcolor{black!10} 1.15    &  \cellcolor{black!10} 13.45   &  \cellcolor{black!10} 1.22 & \cellcolor{black!10} 89.73 &   \cellcolor{black!10}14.98  & &\\ \hline																											
\multirow{4}{*}{Cu$_2$ZnSiS$_4$} & \multirow{4}{*}{3.06}  & 0 &   1                  &  1.53  &   1.61 &  2.67  &  94.58  &  4.03  &  \multirow{4}{*}{N.A.} & \multirow{4}{*}{N.A.}\\
														&        						     & 0 &   $10^{-4}$   & 1.15 &  1.60   &  2.32 & 93.88 &  3.46  & & \\ 	
														&        						     & DFT &  1     &   1.53  &  1.24 &  2.67 & 94.56 &  3.11  & & \\ 
														&        						     & DFT &   $10^{-4}$   & \cellcolor{black!10}  1.15   & \cellcolor{black!10}  1.23   &  \cellcolor{black!10} 2.31 & \cellcolor{black!10} 93.85 & \cellcolor{black!10} 2.66  & & \\ \hline																																	
\end{tabular}%
}
}
\caption{Kesterite Cu$_2$ZnXS$_4$ (X=Sn,Ge,Si)-based solar cell efficiency modelling using the theoretical model proposed by Blank \etal \ \cite{blank2017selection}. Short circuit current density \jsc , open circuit voltage \voc , fill factor $FF$ and cell efficiency $\eta$ values are presented. For each calculation, the optimal absorber layer thickness $d_{\mathrm{opt}}$ has been precalculated and then used as parameter. Results are presented for an internal quantum efficiency $Q_i = 1$ and $Q_i = 10^{-4}$ for experimentally comparable \voc \ values. In order to highlight the impact of the materials reflectivity $R(E)$, the calculation have been performed both for $R(E) =0$ and for $R(E)$ values as obtained using DFT calculations.}
\label{efftable}
\end{table*}
\end{center}

In Table \ref{efftable}, we report the electrical solar cell characteristics for each kesterite material incorporated as the absorber layer with the optimal thickness $d_{\mathrm{opt}}$ and for an internal quantum efficiency $Q_i$. Focusing on the results obtained using DFT-calculated reflectivity $R(E)$ and using an internal quantum efficiency of $Q_i = 10^{-4}$ giving open circuit voltage value comparable to experimental ones \cite{Ratz:2019cs}, solar cell efficiencies of 15.88, 14.98 and 2.66 \% are reported respectively for \czts, \czgs \ and \czss (for an optimal thickness of 1.15 $\mu$m). However, experimentally, lower \jsc \ values around 21.5 mAcm$^{-2}$ and smaller fill factors values between 60 and 65 \% are reported. This observation highlights that the predictions realised with this model corresponds to upper limits. Indeed, nor the materials reflectivity or the absorption of the solar cell upper layers are taken into account, leading to an overestimation of \jsc. Concerning the fill factor, the electrical behaviour of the electrodes is assumed to be ideal. By repeating the calculation with a fixed short circuit current density matching the experimental value, a cell efficiency of 12.29 \% is reported as well as a \voc \ value of 685 mV. This result is in good agreement with the values reported experimentally. 

Using this methodology, we confirmed the interest regarding \czts \ for single-junction solar cell and we highlight a possible efficiency improvement of 10\% which might be achieved by reducing the non-radiative recombination rate. Then, \czgs \ might be interesting as top cell for tandem approaches \cite{vermang2019wide} as this material provides higher bandgap value and interesting cell efficiency, whereas, \czss \ might be interesting for solar cell applications as PV windows.

\section{Conclusion}

In conclusion, we reported direct bandgap values of 1.32, 1.89 and 3.06 eV and absorption coefficients of the order of 10$^{4}$ cm$^{-1}$ for, respectively, \czts, \czgs \ and \czss. Simultaneously a slight increase of the effective mass values is reported following the sequential substitution. Then, using as input data the optical properties of the materials, the solar cell electrical characteristics are predicted based on an improved version of the Shockley-Queisser model. Optimal absorber layer thicknesses between 1.15 and 2.68 $\mu$m are reported and efficiencies of 25.88, 19.94 and 3.11 \% are obtained for the kesterite compounds following the cationic substitution and the induced variation of the materials properties. In addition, using optical results, we highlighted the negative impact of the materials reflectivity on the solar cell characteristics. Using a non-radiative recombination rate giving \voc \ values comparable to actual experimental measurements, we reported a decrease of the solar cell efficiencies to 15.88, 14.98 and 2.66 \% respectively for \czts , \czgs \ and \czss. Pointing out these results as upper limits, by reducing the non-radiative recombination current density, the efficiency of  \czts \ and \czgs \ could be improved respectively by 10 and 4.96 \%, putting forward these kesterite compounds as promising absorber layer materials.

\section*{Conflicts of interest}

There are no conflicts to declare.

\section*{Acknowledgments}

Computational resources have been provided by the Consortium des Équipements de Calcul Intensif (CÉCI), funded by the Fonds de la Recherche Scientifique de Belgique (F.R.S.-FNRS) under Grant No. 2.5020.11 and by the Walloon Region.

\printcredits
\bibliographystyle{model1-num-names}

\bibliography{CZTS_bib}

\begin{thebibliography}{50}
\expandafter\ifx\csname natexlab\endcsname\relax\def\natexlab#1{#1}\fi
\providecommand{\url}[1]{\texttt{#1}}
\providecommand{\href}[2]{#2}
\providecommand{\path}[1]{#1}
\providecommand{\DOIprefix}{doi:}
\providecommand{\ArXivprefix}{arXiv:}
\providecommand{\URLprefix}{URL: }
\providecommand{\Pubmedprefix}{pmid:}
\providecommand{\doi}[1]{\href{http://dx.doi.org/#1}{\path{#1}}}
\providecommand{\Pubmed}[1]{\href{pmid:#1}{\path{#1}}}
\providecommand{\bibinfo}[2]{#2}
\ifx\xfnm\relax \def\xfnm[#1]{\unskip,\space#1}\fi
\bibitem[{Vesborg and Jaramillo(2012)}]{Vesborg:2012gt}
\bibinfo{author}{P.~C.~K. Vesborg}, \bibinfo{author}{T.~F. Jaramillo},
\newblock \bibinfo{title}{{Addressing the terawatt challenge: scalability in
  the supply of chemical elements for renewable energy}},
\newblock \bibinfo{journal}{RSC Advances} \bibinfo{volume}{2}
  (\bibinfo{year}{2012}) \bibinfo{pages}{7933--16}.
\bibitem[{EUC(2017)}]{EUCRM}
\bibinfo{title}{{Communication from the Commision to the European Parliament,
  the Council, the European Economic and Social Committee and the Committee of
  the Regions on the 2017 list of Critical Raw Materials for the EU}},
  \bibinfo{year}{2017}. \URLprefix
  \url{https://eur-lex.europa.eu/legal-content/EN/ALL/?uri=COM:2017:0490:FIN},
  \bibinfo{note}{{Accessed}: 2020-11-18}.
\bibitem[{Green et~al.(2020)Green, Dunlop, Hohl-Ebinger, Yoshita, Kopidakis,
  and Hao}]{Green:2020ga}
\bibinfo{author}{M.~A. Green}, \bibinfo{author}{E.~D. Dunlop},
  \bibinfo{author}{J.~Hohl-Ebinger}, \bibinfo{author}{M.~Yoshita},
  \bibinfo{author}{N.~Kopidakis}, \bibinfo{author}{X.~Hao},
\newblock \bibinfo{title}{{Solar cell efficiency tables (version 56)}},
\newblock \bibinfo{journal}{Progress in Photovoltaics: Research and
  Applications} \bibinfo{volume}{28} (\bibinfo{year}{2020})
  \bibinfo{pages}{629--638}.
\bibitem[{Nakamura et~al.(2019)Nakamura, Yamaguchi, Kimoto, Yasaki, Kato, and
  Sugimoto}]{nakamura2019cd}
\bibinfo{author}{M.~Nakamura}, \bibinfo{author}{K.~Yamaguchi},
  \bibinfo{author}{Y.~Kimoto}, \bibinfo{author}{Y.~Yasaki},
  \bibinfo{author}{T.~Kato}, \bibinfo{author}{H.~Sugimoto},
\newblock \bibinfo{title}{{Cd-free Cu(In, Ga)(Se, S)$_2$ thin-film solar cell
  with record efficiency of 23.35\%}},
\newblock \bibinfo{journal}{IEEE Journal of Photovoltaics} \bibinfo{volume}{9}
  (\bibinfo{year}{2019}) \bibinfo{pages}{1863--1867}.
\bibitem[{Giraldo et~al.(2019)Giraldo, Jehl, Placidi, Izquierdo-Roca,
  P{\'e}rez-Rodr{\'\i}guez, and Saucedo}]{Giraldo:2019iia}
\bibinfo{author}{S.~Giraldo}, \bibinfo{author}{Z.~Jehl},
  \bibinfo{author}{M.~Placidi}, \bibinfo{author}{V.~Izquierdo-Roca},
  \bibinfo{author}{A.~P{\'e}rez-Rodr{\'\i}guez}, \bibinfo{author}{E.~Saucedo},
\newblock \bibinfo{title}{{Progress and perspectives of thin film kesterite
  photovoltaic technology: a critical Review}},
\newblock \bibinfo{journal}{Advanced Materials} \bibinfo{volume}{31}
  (\bibinfo{year}{2019}) \bibinfo{pages}{1806692--18}.
\bibitem[{Wang et~al.(2013)Wang, Winkler, Gunawan, Gokmen, Todorov, Zhu, and
  Mitzi}]{Wang:2013gs}
\bibinfo{author}{W.~Wang}, \bibinfo{author}{M.~T. Winkler},
  \bibinfo{author}{O.~Gunawan}, \bibinfo{author}{T.~Gokmen},
  \bibinfo{author}{T.~K. Todorov}, \bibinfo{author}{Y.~Zhu},
  \bibinfo{author}{D.~B. Mitzi},
\newblock \bibinfo{title}{{Device characteristics of CZTSSe thin-film solar
  cells with 12.6\% efficiency}},
\newblock \bibinfo{journal}{Advanced Energy Materials} \bibinfo{volume}{4}
  (\bibinfo{year}{2013}) \bibinfo{pages}{1301465--5}.
\bibitem[{Yan et~al.(2018)Yan, Huang, Sun, Johnston, Zhang, Sun, Pu, He, Liu,
  Eder, Yang, Cairney, Ekins-Daukes, Hameiri, Stride, Chen, Green, and
  Hao}]{Yan:2018dw}
\bibinfo{author}{C.~Yan}, \bibinfo{author}{J.~Huang}, \bibinfo{author}{K.~Sun},
  \bibinfo{author}{S.~Johnston}, \bibinfo{author}{Y.~Zhang},
  \bibinfo{author}{H.~Sun}, \bibinfo{author}{A.~Pu}, \bibinfo{author}{M.~He},
  \bibinfo{author}{F.~Liu}, \bibinfo{author}{K.~Eder},
  \bibinfo{author}{L.~Yang}, \bibinfo{author}{J.~M. Cairney},
  \bibinfo{author}{N.~J. Ekins-Daukes}, \bibinfo{author}{Z.~Hameiri},
  \bibinfo{author}{J.~A. Stride}, \bibinfo{author}{S.~Chen},
  \bibinfo{author}{M.~A. Green}, \bibinfo{author}{X.~Hao},
\newblock \bibinfo{title}{{Cu$_2$ZnSnS$_4$ solar cells with over 10\% power
  conversion efficiency enabled by heterojunction heat treatment}},
\newblock \bibinfo{journal}{Nature Energy} \bibinfo{volume}{3}
  (\bibinfo{year}{2018}) \bibinfo{pages}{764}.
\bibitem[{Todorov et~al.(2020)Todorov, Hillhouse, Aazou, Sekkat,
  Vigil-Gal{\'a}n, Deshmukh, Agrawal, Bourdais, Vald{\'e}s, Arnou, Mitzi, and
  Dale}]{Todorov:2020ir}
\bibinfo{author}{T.~Todorov}, \bibinfo{author}{H.~W. Hillhouse},
  \bibinfo{author}{S.~Aazou}, \bibinfo{author}{Z.~Sekkat},
  \bibinfo{author}{O.~Vigil-Gal{\'a}n}, \bibinfo{author}{S.~D. Deshmukh},
  \bibinfo{author}{R.~Agrawal}, \bibinfo{author}{S.~Bourdais},
  \bibinfo{author}{M.~Vald{\'e}s}, \bibinfo{author}{P.~Arnou},
  \bibinfo{author}{D.~B. Mitzi}, \bibinfo{author}{P.~J. Dale},
\newblock \bibinfo{title}{{Solution-based synthesis of kesterite thin film
  semiconductors}},
\newblock \bibinfo{journal}{Journal of Physics: Energy} \bibinfo{volume}{2}
  (\bibinfo{year}{2020}) \bibinfo{pages}{012003--22}.
\bibitem[{Ratz et~al.(2019)Ratz, Brammertz, Caballero, Le{\'o}n, Canulescu,
  Schou, G{\"u}tay, Pareek, Taskesen, Kim, Kang, Malerba, Redinger, Saucedo,
  Shin, Tampo, Timmo, Nguyen, and Vermang}]{Ratz:2019cs}
\bibinfo{author}{T.~Ratz}, \bibinfo{author}{G.~Brammertz},
  \bibinfo{author}{R.~Caballero}, \bibinfo{author}{M.~Le{\'o}n},
  \bibinfo{author}{S.~Canulescu}, \bibinfo{author}{J.~Schou},
  \bibinfo{author}{L.~G{\"u}tay}, \bibinfo{author}{D.~Pareek},
  \bibinfo{author}{T.~Taskesen}, \bibinfo{author}{D.-H. Kim},
  \bibinfo{author}{J.~K. Kang}, \bibinfo{author}{C.~Malerba},
  \bibinfo{author}{A.~Redinger}, \bibinfo{author}{E.~Saucedo},
  \bibinfo{author}{B.~Shin}, \bibinfo{author}{H.~Tampo},
  \bibinfo{author}{K.~Timmo}, \bibinfo{author}{N.~D. Nguyen},
  \bibinfo{author}{B.~Vermang},
\newblock \bibinfo{title}{{Physical routes for the synthesis of kesterite}},
\newblock \bibinfo{journal}{Journal of Physics: Energy} \bibinfo{volume}{1}
  (\bibinfo{year}{2019}) \bibinfo{pages}{042003--042024}.
\bibitem[{Grossberg et~al.(2019)Grossberg, Krustok, Hages, Bishop, Gunawan,
  Scheer, Lyam, Hempel, Levcenco, and Unold}]{Grossberg:2019gt}
\bibinfo{author}{M.~Grossberg}, \bibinfo{author}{J.~Krustok},
  \bibinfo{author}{C.~J. Hages}, \bibinfo{author}{D.~M. Bishop},
  \bibinfo{author}{O.~Gunawan}, \bibinfo{author}{R.~Scheer},
  \bibinfo{author}{S.~M. Lyam}, \bibinfo{author}{H.~Hempel},
  \bibinfo{author}{S.~Levcenco}, \bibinfo{author}{T.~Unold},
\newblock \bibinfo{title}{{The electrical and optical properties of
  kesterites}},
\newblock \bibinfo{journal}{Journal of Physics: Energy} \bibinfo{volume}{1}
  (\bibinfo{year}{2019}) \bibinfo{pages}{044002}.
\bibitem[{Platzer-Bj{\"o}rkman et~al.(2019)Platzer-Bj{\"o}rkman, Barreau,
  B{\"a}r, Choubrac, Grenet, Heo, Kubart, Mittiga, S{\'a}nchez, Scragg, Sinha,
  and Valentini}]{PlatzerBjorkman:2019ed}
\bibinfo{author}{C.~Platzer-Bj{\"o}rkman}, \bibinfo{author}{N.~Barreau},
  \bibinfo{author}{M.~B{\"a}r}, \bibinfo{author}{L.~Choubrac},
  \bibinfo{author}{L.~Grenet}, \bibinfo{author}{J.~Heo},
  \bibinfo{author}{T.~Kubart}, \bibinfo{author}{A.~Mittiga},
  \bibinfo{author}{Y.~S{\'a}nchez}, \bibinfo{author}{J.~Scragg},
  \bibinfo{author}{S.~Sinha}, \bibinfo{author}{M.~Valentini},
\newblock \bibinfo{title}{{Back and front contacts in kesterite solar cells:
  state-of-the-art and open questions}},
\newblock \bibinfo{journal}{Journal of Physics: Energy} \bibinfo{volume}{1}
  (\bibinfo{year}{2019}) \bibinfo{pages}{044005--22}.
\bibitem[{Crovetto and Hansen(2017)}]{crovetto2017band}
\bibinfo{author}{A.~Crovetto}, \bibinfo{author}{O.~Hansen},
\newblock \bibinfo{title}{{What is the band alignment of Cu$_2$ZnSn(S,Se)$_4$
  solar cells ?}},
\newblock \bibinfo{journal}{Solar Energy Materials and Solar Cells}
  \bibinfo{volume}{169} (\bibinfo{year}{2017}) \bibinfo{pages}{177--194}.
\bibitem[{Romanyuk et~al.(2019)Romanyuk, Haass, Giraldo, Placidi, Tiwari,
  Fermin, Hao, Xin, Schnabel, Kauk-Kuusik, Pistor, Lie, and
  Wong}]{Romanyuk:2019cq}
\bibinfo{author}{Y.~E. Romanyuk}, \bibinfo{author}{S.~G. Haass},
  \bibinfo{author}{S.~Giraldo}, \bibinfo{author}{M.~Placidi},
  \bibinfo{author}{D.~Tiwari}, \bibinfo{author}{D.~J. Fermin},
  \bibinfo{author}{X.~Hao}, \bibinfo{author}{H.~Xin},
  \bibinfo{author}{T.~Schnabel}, \bibinfo{author}{M.~Kauk-Kuusik},
  \bibinfo{author}{P.~Pistor}, \bibinfo{author}{S.~Lie}, \bibinfo{author}{L.~H.
  Wong},
\newblock \bibinfo{title}{{Doping and alloying of kesterites}},
\newblock \bibinfo{journal}{Journal of Physics: Energy} \bibinfo{volume}{1}
  (\bibinfo{year}{2019}) \bibinfo{pages}{044004--23}.
\bibitem[{Li et~al.(2018)Li, Wang, Li, Zeng, and Zhang}]{Li:2018du}
\bibinfo{author}{J.~Li}, \bibinfo{author}{D.~Wang}, \bibinfo{author}{X.~Li},
  \bibinfo{author}{Y.~Zeng}, \bibinfo{author}{Y.~Zhang},
\newblock \bibinfo{title}{{Cation substitution in Earth-abundant kesterite
  photovoltaic materials}},
\newblock \bibinfo{journal}{Advanced Science} \bibinfo{volume}{5}
  (\bibinfo{year}{2018}) \bibinfo{pages}{1700744--21}.
\bibitem[{Kumar et~al.(2018)Kumar, Madhusudanan, and
  Batabyal}]{kumar2018substitution}
\bibinfo{author}{M.~S. Kumar}, \bibinfo{author}{S.~P. Madhusudanan},
  \bibinfo{author}{S.~K. Batabyal},
\newblock \bibinfo{title}{{Substitution of Zn in Earth-Abundant Cu$_2$ZnSn(S,
  Se)$_4$ based thin film solar cells--A status review}},
\newblock \bibinfo{journal}{Solar Energy Materials and Solar Cells}
  \bibinfo{volume}{185} (\bibinfo{year}{2018}) \bibinfo{pages}{287--299}.
\bibitem[{Tablero(2014)}]{tablero2014electronic}
\bibinfo{author}{C.~Tablero},
\newblock \bibinfo{title}{{Electronic and optical properties of substitutional
  V, Cr and Ir impurities in Cu$_2$ZnSnS$_4$}},
\newblock \bibinfo{journal}{Solar Energy Materials and Solar Cells}
  \bibinfo{volume}{125} (\bibinfo{year}{2014}) \bibinfo{pages}{8--13}.
\bibitem[{Kim et~al.(2016)Kim, Kim, Tampo, Shibata, and Niki}]{Kim:2016jr}
\bibinfo{author}{S.~Kim}, \bibinfo{author}{K.~M. Kim},
  \bibinfo{author}{H.~Tampo}, \bibinfo{author}{H.~Shibata},
  \bibinfo{author}{S.~Niki},
\newblock \bibinfo{title}{{Improvement of voltage deficit of Ge-incorporated
  kesterite solar cell with 12.3\% conversion efficiency}},
\newblock \bibinfo{journal}{Applied Physics Express} \bibinfo{volume}{9}
  (\bibinfo{year}{2016}) \bibinfo{pages}{102301--5}.
\bibitem[{Giraldo et~al.(2018)Giraldo, Saucedo, Neuschitzer, Oliva, Placidi,
  Alcob{\'e}, Izquierdo-Roca, Kim, Tampo, Shibata, P{\'e}rez-Rodr{\'\i}guez,
  and Pistor}]{Giraldo:2018cg}
\bibinfo{author}{S.~Giraldo}, \bibinfo{author}{E.~Saucedo},
  \bibinfo{author}{M.~Neuschitzer}, \bibinfo{author}{F.~Oliva},
  \bibinfo{author}{M.~Placidi}, \bibinfo{author}{X.~Alcob{\'e}},
  \bibinfo{author}{V.~Izquierdo-Roca}, \bibinfo{author}{S.~Kim},
  \bibinfo{author}{H.~Tampo}, \bibinfo{author}{H.~Shibata},
  \bibinfo{author}{A.~P{\'e}rez-Rodr{\'\i}guez}, \bibinfo{author}{P.~Pistor},
\newblock \bibinfo{title}{{How small amounts of Ge modify the formation
  pathways and crystallization of kesterites}},
\newblock \bibinfo{journal}{Energy {\&} Environmental Science}
  \bibinfo{volume}{11} (\bibinfo{year}{2018}) \bibinfo{pages}{582--593}.
\bibitem[{Buffi{\`e}re et~al.(2015)Buffi{\`e}re, ElAnzeery, Oueslati,
  Ben~Messaoud, Brammertz, Meuris, and Poortmans}]{Buffiere:2015dd}
\bibinfo{author}{M.~Buffi{\`e}re}, \bibinfo{author}{H.~ElAnzeery},
  \bibinfo{author}{S.~Oueslati}, \bibinfo{author}{K.~Ben~Messaoud},
  \bibinfo{author}{G.~Brammertz}, \bibinfo{author}{M.~Meuris},
  \bibinfo{author}{J.~Poortmans},
\newblock \bibinfo{title}{{Physical characterization of Cu$_2$ZnGeSe$_4$ thin
  films from annealing of Cu-Zn-Ge precursor layers}},
\newblock \bibinfo{journal}{Thin Solid Films} \bibinfo{volume}{582}
  (\bibinfo{year}{2015}) \bibinfo{pages}{171--175}.
\bibitem[{Choubrac et~al.(2018)Choubrac, Brammertz, Barreau, Arzel, Harel,
  Meuris, and Vermang}]{Choubrac:2018ex}
\bibinfo{author}{L.~Choubrac}, \bibinfo{author}{G.~Brammertz},
  \bibinfo{author}{N.~Barreau}, \bibinfo{author}{L.~Arzel},
  \bibinfo{author}{S.~Harel}, \bibinfo{author}{M.~Meuris},
  \bibinfo{author}{B.~Vermang},
\newblock \bibinfo{title}{{7.6\% CZGSe solar cells thanks to optimized CdS
  chemical bath deposition}},
\newblock \bibinfo{journal}{Physica Status Solidi (a)} \bibinfo{volume}{215}
  (\bibinfo{year}{2018}) \bibinfo{pages}{1800043--9}.
\bibitem[{Vermang et~al.(2019)Vermang, Brammertz, Meuris, Schnabel, Ahlswede,
  Choubrac, Harel, Cardinaud, Arzel, Barreau et~al.}]{vermang2019wide}
\bibinfo{author}{B.~Vermang}, \bibinfo{author}{G.~Brammertz},
  \bibinfo{author}{M.~Meuris}, \bibinfo{author}{T.~Schnabel},
  \bibinfo{author}{E.~Ahlswede}, \bibinfo{author}{L.~Choubrac},
  \bibinfo{author}{S.~Harel}, \bibinfo{author}{C.~Cardinaud},
  \bibinfo{author}{L.~Arzel}, \bibinfo{author}{N.~Barreau}, et~al.,
\newblock \bibinfo{title}{Wide band gap kesterite absorbers for thin film solar
  cells: potential and challenges for their deployment in tandem devices},
\newblock \bibinfo{journal}{Sustainable Energy \& Fuels} \bibinfo{volume}{3}
  (\bibinfo{year}{2019}) \bibinfo{pages}{2246--2259}.
\bibitem[{Khelifi et~al.(2021)Khelifi, Brammertz, Choubrac, Batuk, Yang,
  Meuris, Barreau, Hadermann, Vrielinck, Poelman et~al.}]{khelifi219path}
\bibinfo{author}{S.~Khelifi}, \bibinfo{author}{G.~Brammertz},
  \bibinfo{author}{L.~Choubrac}, \bibinfo{author}{M.~Batuk},
  \bibinfo{author}{S.~Yang}, \bibinfo{author}{M.~Meuris},
  \bibinfo{author}{N.~Barreau}, \bibinfo{author}{J.~Hadermann},
  \bibinfo{author}{H.~Vrielinck}, \bibinfo{author}{D.~Poelman}, et~al.,
\newblock \bibinfo{title}{{The path towards efficient wide band gap thin-film
  kesterite solar cells with transparent back contact for viable tandem
  application}},
\newblock \bibinfo{journal}{Solar Energy Materials and Solar Cells}
  \bibinfo{volume}{219} (\bibinfo{year}{2021}) \bibinfo{pages}{110824}.
\bibitem[{Chen et~al.(2010)Chen, Walsh, Luo, Yang, Gong, and Wei}]{Chen:2010gk}
\bibinfo{author}{S.~Chen}, \bibinfo{author}{A.~Walsh},
  \bibinfo{author}{Y.~Luo}, \bibinfo{author}{J.-H. Yang},
  \bibinfo{author}{X.~G. Gong}, \bibinfo{author}{S.-H. Wei},
\newblock \bibinfo{title}{{Wurtzite-derived polytypes of kesterite and stannite
  quaternary chalcogenide semiconductors}},
\newblock \bibinfo{journal}{Physical Review B} \bibinfo{volume}{82}
  (\bibinfo{year}{2010}) \bibinfo{pages}{Part--8}.
\bibitem[{Khare et~al.(2012)Khare, Himmetoglu, Cococcioni, and
  Aydil}]{Khare:2012gj}
\bibinfo{author}{A.~Khare}, \bibinfo{author}{B.~Himmetoglu},
  \bibinfo{author}{M.~Cococcioni}, \bibinfo{author}{E.~S. Aydil},
\newblock \bibinfo{title}{{First principles calculation of the electronic
  properties and lattice dynamics of Cu$_2$ZnSn(S$_{1-x}$Se$_x$)$_4$}},
\newblock \bibinfo{journal}{Journal of Applied Physics} \bibinfo{volume}{111}
  (\bibinfo{year}{2012}) \bibinfo{pages}{123704--10}.
\bibitem[{Zamulko et~al.(2017)Zamulko, Chen, and Persson}]{Zamulko:2017cy}
\bibinfo{author}{S.~Zamulko}, \bibinfo{author}{R.~Chen},
  \bibinfo{author}{C.~Persson},
\newblock \bibinfo{title}{{Investigation of the structural, optical and
  electronic properties of Cu$_2$Zn(Sn,Si/Ge)(S/Se)$_4$ alloys for solar cell
  applications}},
\newblock \bibinfo{journal}{Physica Status Solidi (b)} \bibinfo{volume}{254}
  (\bibinfo{year}{2017}) \bibinfo{pages}{1700084--5}.
\bibitem[{Liu et~al.(2012)Liu, Chen, Zhai, Xiang, Gong, and Wei}]{Liu:2012dea}
\bibinfo{author}{H.-R. Liu}, \bibinfo{author}{S.~Chen}, \bibinfo{author}{Y.-T.
  Zhai}, \bibinfo{author}{H.~J. Xiang}, \bibinfo{author}{X.~G. Gong},
  \bibinfo{author}{S.-H. Wei},
\newblock \bibinfo{title}{{First-principles study on the effective masses of
  zinc-blend-derived Cu$_2$ZnIVVI$_4$ (IV = Sn, Ge, Si and VI = S,
  Se)}},
\newblock \bibinfo{journal}{Journal of Applied Physics} \bibinfo{volume}{112}
  (\bibinfo{year}{2012}) \bibinfo{pages}{093717--7}.
\bibitem[{Shu et~al.(2013)Shu, Yang, Chen, Huang, Xiang, Gong, and
  Wei}]{Shu:2013ed}
\bibinfo{author}{Q.~Shu}, \bibinfo{author}{J.-H. Yang},
  \bibinfo{author}{S.~Chen}, \bibinfo{author}{B.~Huang},
  \bibinfo{author}{H.~Xiang}, \bibinfo{author}{X.-G. Gong},
  \bibinfo{author}{S.-H. Wei},
\newblock \bibinfo{title}{{Cu$_2$Zn(Sn,Ge)Se$_4$ and Cu$_2$Zn(Sn,Si)Se$_4$
  alloys as photovoltaic materials: Structural and electronic properties}},
\newblock \bibinfo{journal}{Physical Review B} \bibinfo{volume}{87}
  (\bibinfo{year}{2013}) \bibinfo{pages}{364--6}.
\bibitem[{Kim et~al.(2018)Kim, Park, and Walsh}]{Kim:2018jd}
\bibinfo{author}{S.~Kim}, \bibinfo{author}{J.-S. Park},
  \bibinfo{author}{A.~Walsh},
\newblock \bibinfo{title}{{Identification of killer defects in kesterite
  thin-film solar cells}},
\newblock \bibinfo{journal}{ACS Energy Letters} \bibinfo{volume}{3}
  (\bibinfo{year}{2018}) \bibinfo{pages}{496--500}.
\bibitem[{Chen et~al.(2013)Chen, Walsh, Gong, and Wei}]{Chen:2013cna}
\bibinfo{author}{S.~Chen}, \bibinfo{author}{A.~Walsh}, \bibinfo{author}{X.-G.
  Gong}, \bibinfo{author}{S.-H. Wei},
\newblock \bibinfo{title}{{Classification of lattice defects in the kesterite
  Cu$_2$ZnSnS$_4$ and Cu$_2$ZnSnSe$_4$ earth-abundant solar cell absorbers}},
\newblock \bibinfo{journal}{Advanced Materials} \bibinfo{volume}{25}
  (\bibinfo{year}{2013}) \bibinfo{pages}{1522--1539}.
\bibitem[{Heyd et~al.(2003)Heyd, Scuseria, and Ernzerhof}]{heyd2003hybrid}
\bibinfo{author}{J.~Heyd}, \bibinfo{author}{G.~E. Scuseria},
  \bibinfo{author}{M.~Ernzerhof},
\newblock \bibinfo{title}{{Hybrid functionals based on a screened Coulomb
  potential}},
\newblock \bibinfo{journal}{The Journal of Chemical Physics}
  \bibinfo{volume}{118} (\bibinfo{year}{2003}) \bibinfo{pages}{8207--8215}.
\bibitem[{Blank et~al.(2017)Blank, Kirchartz, Lany, and
  Rau}]{blank2017selection}
\bibinfo{author}{B.~Blank}, \bibinfo{author}{T.~Kirchartz},
  \bibinfo{author}{S.~Lany}, \bibinfo{author}{U.~Rau},
\newblock \bibinfo{title}{{Selection metric for photovoltaic materials
  screening based on detailed-balance analysis}},
\newblock \bibinfo{journal}{Physical Review Applied} \bibinfo{volume}{8}
  (\bibinfo{year}{2017}) \bibinfo{pages}{024032}.
\bibitem[{Kresse and Furthm{\"u}ller(1996)}]{kresse1996efficiency}
\bibinfo{author}{G.~Kresse}, \bibinfo{author}{J.~Furthm{\"u}ller},
\newblock \bibinfo{title}{{Efficiency of ab-initio total energy calculations
  for metals and semiconductors using a plane-wave basis set}},
\newblock \bibinfo{journal}{Computational Materials Science}
  \bibinfo{volume}{6} (\bibinfo{year}{1996}) \bibinfo{pages}{15--50}.
\bibitem[{Kresse and Joubert(1999)}]{kresse1999ultrasoft}
\bibinfo{author}{G.~Kresse}, \bibinfo{author}{D.~Joubert},
\newblock \bibinfo{title}{{From ultrasoft pseudopotentials to the projector
  augmented-wave method}},
\newblock \bibinfo{journal}{Physical Review B} \bibinfo{volume}{59}
  (\bibinfo{year}{1999}) \bibinfo{pages}{1758}.
\bibitem[{Perdew et~al.(1996)Perdew, Burke, and
  Ernzerhof}]{perdew1996generalized}
\bibinfo{author}{J.~P. Perdew}, \bibinfo{author}{K.~Burke},
  \bibinfo{author}{M.~Ernzerhof},
\newblock \bibinfo{title}{{Generalized gradient approximation made simple}},
\newblock \bibinfo{journal}{Physical Review Letters} \bibinfo{volume}{77}
  (\bibinfo{year}{1996}) \bibinfo{pages}{3865}.
\bibitem[{Sun et~al.(2015)Sun, Ruzsinszky, and Perdew}]{sun2015strongly}
\bibinfo{author}{J.~Sun}, \bibinfo{author}{A.~Ruzsinszky},
  \bibinfo{author}{J.~P. Perdew},
\newblock \bibinfo{title}{{Strongly constrained and appropriately normed
  semilocal density functional}},
\newblock \bibinfo{journal}{Physical Review Letters} \bibinfo{volume}{115}
  (\bibinfo{year}{2015}) \bibinfo{pages}{036402}.
\bibitem[{Sun et~al.(2016)Sun, Remsing, Zhang, Sun, Ruzsinszky, Peng, Yang,
  Paul, Waghmare, Wu et~al.}]{sun2016accurate}
\bibinfo{author}{J.~Sun}, \bibinfo{author}{R.~C. Remsing},
  \bibinfo{author}{Y.~Zhang}, \bibinfo{author}{Z.~Sun},
  \bibinfo{author}{A.~Ruzsinszky}, \bibinfo{author}{H.~Peng},
  \bibinfo{author}{Z.~Yang}, \bibinfo{author}{A.~Paul},
  \bibinfo{author}{U.~Waghmare}, \bibinfo{author}{X.~Wu}, et~al.,
\newblock \bibinfo{title}{{Accurate first-principles structures and energies of
  diversely bonded systems from an efficient density functional}},
\newblock \bibinfo{journal}{Nature Chemistry} \bibinfo{volume}{8}
  (\bibinfo{year}{2016}) \bibinfo{pages}{831}.
\bibitem[{Heyd et~al.(2005)Heyd, Peralta, Scuseria, and
  Martin}]{heyd2005energy}
\bibinfo{author}{J.~Heyd}, \bibinfo{author}{J.~E. Peralta},
  \bibinfo{author}{G.~E. Scuseria}, \bibinfo{author}{R.~L. Martin},
\newblock \bibinfo{title}{{Energy band gaps and lattice parameters evaluated
  with the Heyd-Scuseria-Ernzerhof screened hybrid functional}},
\newblock \bibinfo{journal}{The Journal of Chemical Physics}
  \bibinfo{volume}{123} (\bibinfo{year}{2005}) \bibinfo{pages}{174101}.
\bibitem[{Guo et~al.(2009)Guo, Hillhouse, and Agrawal}]{guo2009synthesis}
\bibinfo{author}{Q.~Guo}, \bibinfo{author}{H.~W. Hillhouse},
  \bibinfo{author}{R.~Agrawal},
\newblock \bibinfo{title}{{Synthesis of Cu$_2$ZnSnS$_4$ nanocrystal ink and its
  use for solar cells}},
\newblock \bibinfo{journal}{Journal of the American Chemical Society}
  \bibinfo{volume}{131} (\bibinfo{year}{2009}) \bibinfo{pages}{11672--11673}.
\bibitem[{Levcenko et~al.(2012)Levcenko, Tezlevan, Arushanov, Schorr, and
  Unold}]{levcenko2012free}
\bibinfo{author}{S.~Levcenko}, \bibinfo{author}{V.~Tezlevan},
  \bibinfo{author}{E.~Arushanov}, \bibinfo{author}{S.~Schorr},
  \bibinfo{author}{T.~Unold},
\newblock \bibinfo{title}{{Free-to-bound recombination in near stoichiometric
  Cu$_2$ZnSnS$_4$ single crystals}},
\newblock \bibinfo{journal}{Physical Review B} \bibinfo{volume}{86}
  (\bibinfo{year}{2012}) \bibinfo{pages}{045206}.
\bibitem[{Lisunov et~al.(2013)Lisunov, Guk, Nateprov, Levcenko, Tezlevan, and
  Arushanov}]{lisunov2013features}
\bibinfo{author}{K.~Lisunov}, \bibinfo{author}{M.~Guk},
  \bibinfo{author}{A.~Nateprov}, \bibinfo{author}{S.~Levcenko},
  \bibinfo{author}{V.~Tezlevan}, \bibinfo{author}{E.~Arushanov},
\newblock \bibinfo{title}{{Features of the acceptor band and properties of
  localized carriers from studies of the variable-range hopping conduction in
  single crystals of p-Cu$_2$ZnSnS$_4$}},
\newblock \bibinfo{journal}{Solar Energy Materials and Solar Cells}
  \bibinfo{volume}{112} (\bibinfo{year}{2013}) \bibinfo{pages}{127--133}.
\bibitem[{Dun et~al.(2014)Dun, Holzwarth, Li, Huang, and Carroll}]{Dun:2014dt}
\bibinfo{author}{C.~Dun}, \bibinfo{author}{N.~A.~W. Holzwarth},
  \bibinfo{author}{Y.~Li}, \bibinfo{author}{W.~Huang}, \bibinfo{author}{D.~L.
  Carroll},
\newblock \bibinfo{title}{{Cu$_2$ZnSn(S$_x$O$_{(1-x)}$)$_4$ and
  Cu$_2$ZnSn(S$_x$Se$_{(1-x)}$)$_4$: First principles simulations of optimal
  alloy configurations and their energies}},
\newblock \bibinfo{journal}{Journal of Applied Physics} \bibinfo{volume}{115}
  (\bibinfo{year}{2014}) \bibinfo{pages}{193513--13}.
\bibitem[{Walsh et~al.(2012)Walsh, Chen, Wei, and Gong}]{walsh2012kesterite}
\bibinfo{author}{A.~Walsh}, \bibinfo{author}{S.~Chen}, \bibinfo{author}{S.-H.
  Wei}, \bibinfo{author}{X.-G. Gong},
\newblock \bibinfo{title}{{Kesterite thin-film solar cells: Advances in
  materials modelling of Cu$_2$ZnSnS$_4$}},
\newblock \bibinfo{journal}{Advanced Energy Materials} \bibinfo{volume}{2}
  (\bibinfo{year}{2012}) \bibinfo{pages}{400--409}.
\bibitem[{Khadka and Kim(2013)}]{khadka2013study}
\bibinfo{author}{D.~B. Khadka}, \bibinfo{author}{J.~Kim},
\newblock \bibinfo{title}{{Study of structural and optical properties of
  kesterite Cu$_2$ZnGeX$_4$ (X= S, Se) thin films synthesized by chemical spray
  pyrolysis}},
\newblock \bibinfo{journal}{CrystEngComm} \bibinfo{volume}{15}
  (\bibinfo{year}{2013}) \bibinfo{pages}{10500--10509}.
\bibitem[{Hamdi et~al.(2014)Hamdi, Lafond, Guillot-Deudon, Hlel, Gargouri, and
  Jobic}]{hamdi2014crystal}
\bibinfo{author}{M.~Hamdi}, \bibinfo{author}{A.~Lafond},
  \bibinfo{author}{C.~Guillot-Deudon}, \bibinfo{author}{F.~Hlel},
  \bibinfo{author}{M.~Gargouri}, \bibinfo{author}{S.~Jobic},
\newblock \bibinfo{title}{{Crystal chemistry and optical investigations of the
  Cu$_2$Zn(Sn,Si)S$_4$ series for photovoltaic applications}},
\newblock \bibinfo{journal}{Journal of Solid State Chemistry}
  \bibinfo{volume}{220} (\bibinfo{year}{2014}) \bibinfo{pages}{232--237}.
\bibitem[{Levcenco et~al.(2011)Levcenco, Dumcenco, Huang, Arushanov, Tezlevan,
  Tiong, and Du}]{levcenco2011polarization}
\bibinfo{author}{S.~Levcenco}, \bibinfo{author}{D.~Dumcenco},
  \bibinfo{author}{Y.~Huang}, \bibinfo{author}{E.~Arushanov},
  \bibinfo{author}{V.~Tezlevan}, \bibinfo{author}{K.~Tiong},
  \bibinfo{author}{C.~Du},
\newblock \bibinfo{title}{{Polarization-dependent electrolyte
  electroreflectance study of Cu$_2$ZnSiS$_4$ and Cu$_2$ZnSiSe$_4$ single
  crystals}},
\newblock \bibinfo{journal}{Journal of Alloys and Compounds}
  \bibinfo{volume}{509} (\bibinfo{year}{2011}) \bibinfo{pages}{7105--7108}.
\bibitem[{Slater(1964)}]{slater1964atomic}
\bibinfo{author}{J.~C. Slater},
\newblock \bibinfo{title}{{Atomic radii in crystals}},
\newblock \bibinfo{journal}{The Journal of Chemical Physics}
  \bibinfo{volume}{41} (\bibinfo{year}{1964}) \bibinfo{pages}{3199--3204}.
\bibitem[{Zhang et~al.(2012)Zhang, Sun, Zhang, Yuan, Huang, and
  Zhang}]{zhang2012structural}
\bibinfo{author}{Y.~Zhang}, \bibinfo{author}{X.~Sun},
  \bibinfo{author}{P.~Zhang}, \bibinfo{author}{X.~Yuan},
  \bibinfo{author}{F.~Huang}, \bibinfo{author}{W.~Zhang},
\newblock \bibinfo{title}{{Structural properties and quasiparticle band
  structures of Cu-based quaternary semiconductors for photovoltaic
  applications}},
\newblock \bibinfo{journal}{Journal of Applied Physics} \bibinfo{volume}{111}
  (\bibinfo{year}{2012}) \bibinfo{pages}{063709}.
\bibitem[{Vishwakarma et~al.(2018)Vishwakarma, Varandani, Shivaprasad, and
  Mehta}]{vishwakarma2018structural}
\bibinfo{author}{M.~Vishwakarma}, \bibinfo{author}{D.~Varandani},
  \bibinfo{author}{S.~Shivaprasad}, \bibinfo{author}{B.~Mehta},
\newblock \bibinfo{title}{{Structural, optical, electrical properties and
  energy band diagram of Cu$_2$ZnSiS$_4$ thin films}},
\newblock \bibinfo{journal}{Solar Energy Materials and Solar Cells}
  \bibinfo{volume}{174} (\bibinfo{year}{2018}) \bibinfo{pages}{577--583}.
\bibitem[{Paier et~al.(2009)Paier, Asahi, Nagoya, and Kresse}]{paier2009cu}
\bibinfo{author}{J.~Paier}, \bibinfo{author}{R.~Asahi},
  \bibinfo{author}{A.~Nagoya}, \bibinfo{author}{G.~Kresse},
\newblock \bibinfo{title}{{Cu$_2$ZnSnS$_4$ as a potential photovoltaic
  material: a hybrid Hartree-Fock density functional theory study}},
\newblock \bibinfo{journal}{Physical Review B} \bibinfo{volume}{79}
  (\bibinfo{year}{2009}) \bibinfo{pages}{115126}.
\bibitem[{Kim et~al.(2020)Kim, M{\'a}rquez, Unold, and Walsh}]{kim2020upper}
\bibinfo{author}{S.~Kim}, \bibinfo{author}{J.~A. M{\'a}rquez},
  \bibinfo{author}{T.~Unold}, \bibinfo{author}{A.~Walsh},
\newblock \bibinfo{title}{{Upper limit to the photovoltaic efficiency of
  imperfect crystals from first principles}},
\newblock \bibinfo{journal}{Energy \& Environmental Science}
  \bibinfo{volume}{13} (\bibinfo{year}{2020}) \bibinfo{pages}{1481--1491}.

\end{thebibliography}

\end{document}